\documentclass[draft,jgrga]{agutex2015}

\usepackage{amsmath}			
\usepackage{amssymb}
\usepackage{lscape}			
\usepackage{longtable}
\usepackage{rotating}
\usepackage[table]{xcolor}
\usepackage{multirow}
\usepackage{color, soul} 

\setkeys{Gin}{draft=false}
\authorrunninghead{H\"ORST}

\titlerunninghead{TITAN'S ATMOSPHERE}

\begin{document}

%
%

\title{Titan's Atmosphere and Climate}
%
%

%
%



 \authors{S. M. H\"orst\altaffilmark{1}}

\altaffiltext{1}{Department of Earth and Planetary Sciences,
Johns Hopkins University, Baltimore, MD, USA.}





%
%


\keypoints{\item Titan has the most complex atmospheric chemistry in the solar system. \item Titan's atmosphere and surface share a unique connection. \item Titan provides the opportunity to test our understanding of many planetary processes}


%
%


\begin{abstract}

Titan is the only moon with a substantial atmosphere, the only other thick N$_{2}$ atmosphere besides Earth's, the site of extraordinarily complex atmospheric chemistry that far surpasses any other solar system atmosphere, and the only other solar system body with stable liquid currently on its surface. The connection between Titan's surface and atmosphere is also unique in our Solar system; atmospheric chemistry produces materials that are deposited on the surface and subsequently altered by surface-atmosphere interactions such as aeolian and fluvial processes resulting in the formation of extensive dune fields and expansive lakes and seas. Titan's atmosphere is favorable for organic haze formation, which combined with the presence of some oxygen bearing molecules indicates that Titan's atmosphere may produce molecules of prebiotic interest. The combination of organics and liquid, in the form of water in a subsurface ocean and methane/ethane in the surface lakes and seas, means that Titan may be the ideal place in the solar system to test ideas about habitability, prebiotic chemistry, and the ubiquity and diversity of life in the Universe. The Cassini-Huygens mission to the Saturn system has provided a wealth of new information allowing for study of Titan as a complex system. Here I review our current understanding of Titan's atmosphere and climate forged from the powerful combination of Earth-based observations, remote sensing and \emph{in situ} spacecraft measurements, laboratory experiments, and models. I conclude with some of our remaining unanswered questions as the incredible era of exploration with Cassini-Huygens comes to an end.
\end{abstract}

%
%

%

\begin{article}

%
%

\section{Introduction}

Titan is unique in our solar system: it is the only moon with a substantial atmosphere, the only other thick N$_{2}$ atmosphere besides that of Earth, the site of extraordinarily complex atmospheric chemistry that far surpasses any other solar system atmosphere, and the only other solar system body that currently possesses stable liquid on its surface. Titan's mildly reducing atmosphere is favorable for organic haze formation and the presence of some oxygen bearing molecules suggests that molecules of prebiotic interest may form in its atmosphere \citep{Horst:2012}. The combination of liquid and organics means that Titan may be the ideal place in the solar system to test ideas about habitability, prebiotic chemistry, and the ubiquity and diversity of life in the Universe \citep{Lunine:2009b}.

The possible existence of an atmosphere around Titan was first suggested in 1908 in a paper titled ``Observations des satellites principaux de Jupiter et de Titan'' by Jos\'e Comas Sol\`a. He believed he observed limb darkening, which is indicative of an atmosphere \citep{Sola:1908}. Gerard Kuiper's discovery of methane (CH$_{4}$) around Titan provided conclusive evidence that Titan possesses an atmosphere \citep{Kuiper:1944}. 

During the 1970s, analyses of ground-based infrared spectra began to provide constraints on the temperature structure \citep{Morrison:1972, Danielson:1973} and composition \citep{Trafton:1972, Trafton:1972b, Gillett:1973, Gillett:1975} of Titan's atmosphere. The detection of a greenhouse effect \citep{Morrison:1972} coupled with detailed investigations of the spectral features of CH$_{4}$ \citep{Trafton:1972b, Lutz:1976} led to the conclusion that CH$_{4}$ might be only a minor constituent in Titan's atmosphere and that other absorbers, such as C$_{2}$H$_{6}$ or some type of dust particle, might be present \citep{Danielson:1973}. The presence of N$_{2}$ in Titan's atmosphere was first suggested by \citet{Lewis:1971} based on the idea that Titan accreted NH$_{3}$ during formation, which later photolyzed to form N$_{2}$. Photochemical models predicted significant abundances of C$_{2}$H$_{2}$ and C$_{2}$H$_{6}$ \citep{Strobel:1974, Allen:1980}, although only a few models included N$_{2}$ and nitrogen chemistry \citep{Atreya:1978}. The encounter of Pioneer 11 with Titan in 1979 confirmed the presence of an aerosol absorber in the atmosphere \citep{Tomasko:1980} and provided a lower limit of 1.37 g/cm$^{2}$ on Titan's density \citep{Smith:1980}. The encounters of Voyager 1 and 2 with Titan in the early 1980s confirmed that Titan possesses a substantial N$_{2}$ atmosphere, with a surface pressure 1.5 times that of Earth and a surface temperature of 94 K \citep{Broadfoot:1981, Lindal:1983, Hunten:1984}. Infrared spectra taken by the Voyager spacecraft revealed the presence of a variety of organic molecules (C$_{2}$H$_{2}$, C$_{2}$H$_{4}$, C$_{2}$H$_{6}$, C$_{3}$H$_{8}$, CH$_{3}$C$_{2}$H, C$_{4}$H$_{2}$, etc.) \citep{Hanel:1981, Kunde:1981} and provided the first glimpse of the complexity of the chemical and physical processes occurring in Titan's atmosphere.

Despite providing an enormous amount of new information about Titan's atmosphere, Titan's thick photochemical haze and abundant methane prevented the instruments carried by Pioneer 11 and the Voyager spacecraft from seeing Titan's surface. Images beamed back to Earth as the spacecraft sped through the Saturnian system showed only a featureless orange ball, providing no hint of the incredible landscape below. It remained shrouded for another 23 years until the arrival of the Cassini-Huygens mission in 2004. 

The arrival of the Cassini-Huygens mission to the Saturn system ushered in a new era in the study of Titan. Carrying a variety of instruments capable of remote sensing and \emph{in situ} investigations of Titan's atmosphere and surface, the Cassini Orbiter and the Huygens Probe have provided a wealth of new information about Titan and have finally allowed humankind to see the surface. Perhaps more so than anywhere else in the solar system, Titan's atmosphere and surface are intimately linked. As discussed in detail below, the organic material transported to form Titan's extensive sand seas was formed initially by chemical and physical processes in the atmosphere, the liquids that carve out dendritic channels like those seen at the Huygens landing site cycle between the atmosphere and the surface (in the case of methane) or are produced by chemistry in the atmosphere (in the case of ethane, propane, hydrogen cyanide, etc). Long term climate cycles result in the observed asymmetry of lakes and seas on the surface \citep{Aharonson:2009}. Cryovolcanism may resupply methane and other gases to the atmosphere. 

Here I review our current understanding of Titan's atmosphere and climate forged from the powerful combination of Earth-based observations, remote sensing and \emph{in situ} spacecraft measurements, laboratory experiments, and models. I conclude with a discussion of some of our remaining unanswered questions as the incredible era of exploration with Cassini-Huygens comes to an end.

\section{Titan's atmospheric structure and composition}

The two main constituents of Titan's atmosphere are molecular nitrogen (N$_{2}$) and methane (CH$_{4}$). Titan receives about 1\% of the solar flux that reaches Earth. Of the flux incident at the top of Titan's atmosphere, only 10\% reaches the surface (compared to 57\% for Earth) (see e.g., \citet{Griffith:2012, Read:2015}). Titan is therefore much colder than Earth, with an effective temperature of $\sim$82 K. The combination of the greenhouse effect provided by CH$_{4}$ and collision induced absorption (N$_{2}$-N$_{2}$, N$_{2}$-CH$_{4}$, N$_{2}$-H$_{2}$) and the anti-greenhouse from the stratospheric haze layer \citep{McKay:1991} results in a surface temperature of approximately 94 K \citep{Lindal:1983, Fulchignoni:2005, Schinder:2011}. The pressure at Titan's surface is $\sim$1.5 bar \citep{Lindal:1983, Fulchignoni:2005} resulting in surface conditions that are near the triple point of methane, much like water on Earth, which allows for liquid methane on the surface and gaseous methane in the atmosphere. The fluvial and aeolian features on the surface, discussed in Section \ref{Sect:surface} indicate that Titan has an active ``hydrological'' cycle that is, in many ways, both very similar to and very different from that of Earth. In addition to the features observed on the surface, large storms occasionally erupt in Titan's atmosphere and are presumably responsible for many of the fluvial features.  

\subsection{Temperature structure}

\begin{figure}
{\includegraphics[width=0.9\textwidth]{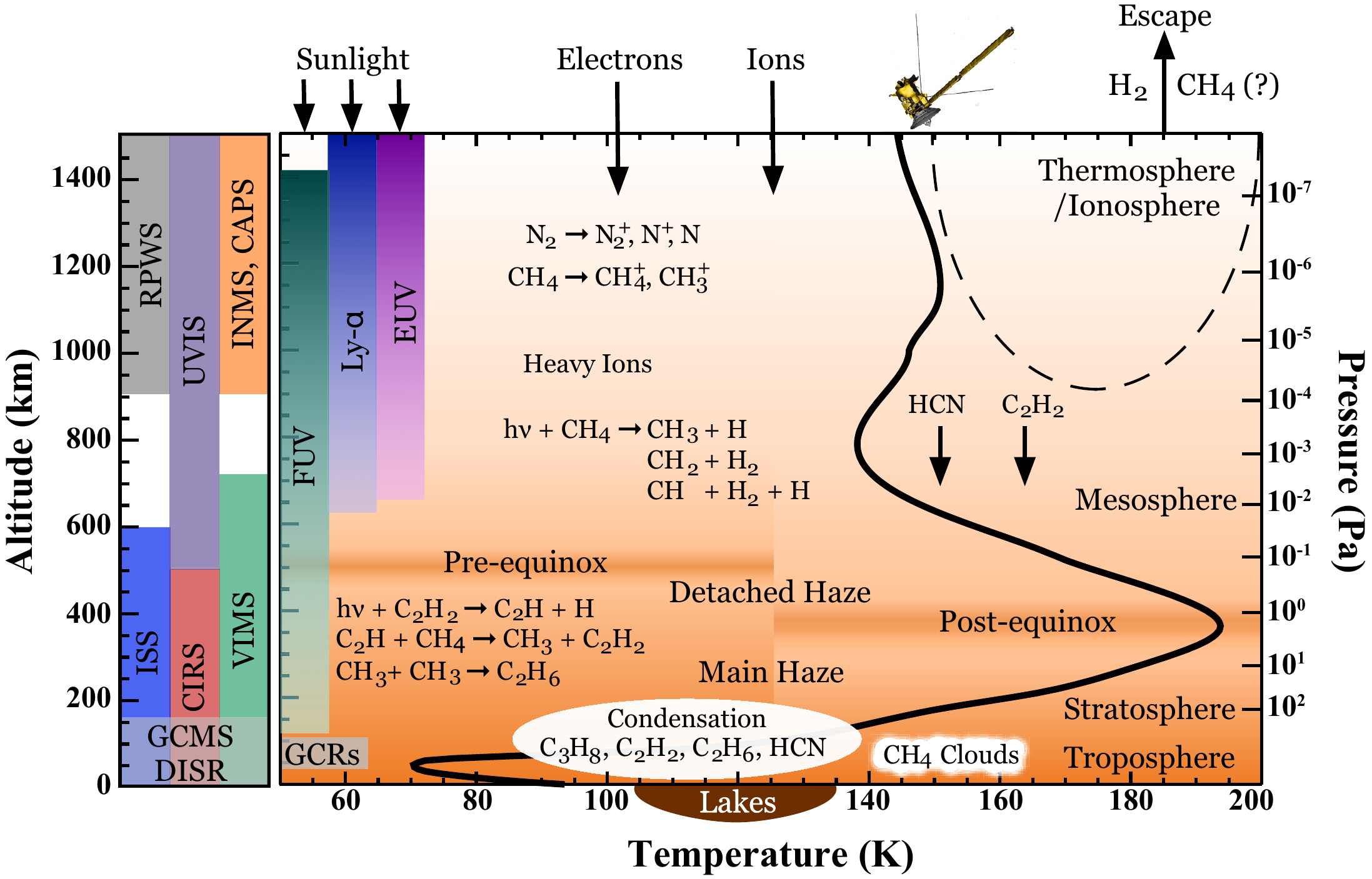}}
\caption{Shown here is a representative temperature profile for Titan's atmosphere, some of the major chemical processes, and the approximate altitude coverage of the instruments carried by Cassini-Huygens.}
\label{fig:titanatm}
\end{figure}

The vertical temperature structure in Titan's atmosphere (see Figure \ref{fig:titanatm}) is most analogous to that of Earth, with well defined tropo-, strato-, meso-, and thermospheres (see e.g., \citet{SmithG:1982, Lindal:1983, Hubbard:1990, Sicardy:1990, Fulchignoni:2005}). Although Titan's atmosphere is colder than Earth's, Titan's atmosphere is more extended, with scale heights of 15 to 50 km compared to 5 to 8 km on Earth due to Titan's lower gravity \citep{Flasar:2014}. Finer scale structure in temperature measurements attributed to atmospheric waves has also been observed throughout the atmosphere (see e.g., \citet{Sicardy:1999, Fulchignoni:2005, MullerWodarg:2006, Sicardy:2006, Strobel:2006, Zalucha:2007, Aboudan:2008, Koskinen:2011, Lorenz:2014d}). Temperatures in the thermosphere are similar to those in the mesosphere, despite the significant EUV heating rates in the thermosphere, due to efficient cooling by HCN rotational lines \citep{Yelle:1991}.

The thermal structure of the upper atmosphere is more variable and complex than expected and there are still a number of outstanding questions regarding its variability. On average the night side is warmer than the day side \citep{MullerWodarg:2006, delahaye:2007, Cui:2009, Westlake:2011, Snowden:2013}. There does not seem to be any correlation between temperature and latitude indicating that solar input does not control the temperature \citep{Yelle:2014}. Additionally, the highest temperatures are seen when Titan is inside Saturn's magnetosphere, while the coldest temperatures are measured when Titan is outside the magnetosphere \citep{Westlake:2011, Snowden:2013}.

The temperature profile measured by the Huygens Atmospheric Structure Instrument (HASI) finally constrained the location of Titan's planetary boundary layer (PBL), which is important for understanding surface-atmosphere interactions. Analyses of the HASI data find that the PBL is located at 300 m \citep{Fulchignoni:2005, Tokano:2006b}, which was incompatible with analyses of other temperature measurements \citep{Lindal:1983, Schinder:2011} and dune spacing \citep{Lorenz:2010} that find the PBL is located at 2-3 km. The General Circulation Model (GCM) of \citet{Charnay:2012} showed that Titan should have two PBLs, a seasonal one located at 2 km and a diurnal one that reaches up to 800 m over the course of a day and the Huygens' measurement at 300 m is consistent with the mid-morning arrival of the Huygens Probe. The presence of diurnal and seasonal boundary layers indicates that Titan's troposphere is more dynamic than previously thought. 

\subsection{Methane abundance \label{sec:methane}}

 Deep in the troposphere and at the surface, the CH$_{4}$ mixing ratio is obtained only by \emph{in situ} measurements and the Gas Chromatograph Mass Spectrometer (GCMS) carried by Huygens found the methane mixing ratio at the surface to be 5.65 $\pm$ 0.18\% \citep{Niemann:2010}. This value is consistent with estimates made from spectra taken by the Descent Imager/Spectral Radiometer (DISR) in the last 25 m of the descent \citep{Schroder:2008, Jacquemart:2008}. The mixing ratio is constant within the error bars until about 7 km where it decreases with altitude until near 45 km where it is constant throughout the rest of the GCMS measurements (up to 140 km). The decrease is a result of condensation and the temperature at the tropopause limits the methane mixing ratio in the stratosphere. Unfortunately, this profile is limited to a single location at a single point in time (10.3$^{\circ}$S 167.7$^{\circ}$W, 14 January 2005), and there is some evidence of tropospheric methane variations of $\sim$10-40\% \citep{Tokano:2014, Adamkovics:2016b}. The average stratospheric value from GCMS is 1.48 $\pm$ 0.09\%, which is consistent with initial determinations of the stratospheric methane mixing ratio from the Composite Infrared Spectrometer (CIRS) (1.6 $\pm$ 0.5\%, \citet{Flasar:2005}) and DISR \citep{Bezard:2014}. However, measurements from the Cassini Visible and Infrared Mapping Spectrometer (VIMS) (1.28 $\pm$ 0.06\%, \citet{Maltagliati:2015}) and Herschel SPIRE (1.33 $\pm$ 0.07\%, \citet{Courtin:2011}) and PACS (1.29 $\pm$ 0.03\%, \citet{Rengel:2014}) find lower stratospheric values. Recent analyses of Cassini CIRS measurements find variation with latitude from 1\% at low latitudes to 1.5\% at polar latitudes, indicating that the methane mixing ratio in the stratosphere may be more complicated than previously thought \citep{Lellouch:2014}. In the mesosphere composition is measured by stellar/solar UV occultation or UV dayglow and the CH$_{4}$ profile was measured by Voyager UVS (Ultraviolet Spectrometer) \citep{Smith:1982, Vervack:2004} and the Cassini Ultraviolet Imaging Spectrograph (UVIS) \citep{Shemansky:2005, Koskinen:2011, Stevens:2015}. The mesospheric methane abundance appears to be relatively constant with altitude until it begins diffusively separating around 900 km \citep{Koskinen:2011, Capalbo:2013, Stevens:2015}. From 950 to 1500 km, the CH$_{4}$ mixing ratio is measured by the Ion and Neutral Mass Spectrometer (INMS) \citep{MullerWodarg:2008, Magee:2009, Cui:2009} and varies from 1.31\% at 981 km to 3\% at 1150 km. These results are also consistent with UVIS EUV occultation measurements that probe up to 1600 km \citep{Kammer:2013, Capalbo:2015}.

\subsection{Voyager and Earth-based composition measurements}
The Infrared Radiometer Interferometer and Spectrometer (IRIS) carried by Voyager first revealed the complexity of Titan's atmospheric chemistry with the measurements of methane (CH$_{4}$), molecular hydrogen (H$_{2}$), ethane (C$_{2}$H$_{6}$), acetylene (C$_{2}$H$_{2}$), ethylene (C$_{2}$H$_{4}$), hydrogen cyanide (HCN), methylacetylene (C$_{3}$H$_{4}$), propane (C$_{3}$H$_{8}$), diacetylene (C$_{4}$H$_{2}$), cyanoacetylene (HC$_{3}$N), cyanogen (C$_{2}$N$_{2}$), and carbon dioxide (CO$_{2}$) \citep{Hanel:1981, Kunde:1981, Samuelson:1981,Samuelson:1983}. In the decades between the Voyager encounters and Cassini-Huygens' arrival in the Saturn system, carbon monoxide (CO) (\citet{Lutz:1983}, Mayall Telescope Kitt Peak), acetonitrile (CH$_{3}$CN) (\citet{Bezard:1993, Marten:2002}, IRAM 30 m), water (H$_{2}$O) (\citet{Coustenis:1998}, ISO), and benzene (C$_{6}$H$_{6}$) (\citep{Coustenis:2003}, ISO) were discovered using Earth-based telescopes. Earth-based telescopes have continued to discover new molecules in Titan's atmosphere during the Cassini-Huygens mission; Herschel measurements revealed the presence of hydrogen isocyanide (HNC) \citep{Moreno:2011} and ethyl cyanide (C$_{2}$H$_{5}$CN) was discovered in measurements from the Atacama Large Millimeter Array (ALMA) \citep{Cordiner:2015}, demonstrating the capabilities and importance of Earth-based observations. 

\subsection{Composition measurements from Cassini-Huygens}
Cassini-Huygens carries a powerful combination of remote sensing and \emph{in situ} instruments allowing for detection and measurement of gas phase species and characterization of aerosol particles from the troposphere to the ionosphere. CIRS has provided the temporal and spatial measurements of Titan's stratosphere necessary to understand the small molecule chemistry and constrain dynamics and seasonal change. CIRS also discovered propene (C$_{3}$H$_{6}$) \citep{Nixon:2013b}. UVIS senses the previously unexplored mesosphere providing composition and temperature information necessary to complete our understanding of Titan's atmospheric structure. As discussed in Section \ref{sec:age}, GCMS and INMS are particularly important because they measure noble gases and $^{14}$N/$^{15}$N in nitrogen, which are powerful tools for understanding the origin and evolution of atmospheres but cannot be detected from remote sensing. Additionally, the GCMS carried to the surface and INMS and CAPS carried by the orbiter allow us to measure regions of the atmosphere that cannot be accessed with remote sensing due to density. Numerous previously undetected species have been identified in measurements obtained by the Ion and Neutral Mass spectrometer in Titan's thermosphere \citep{Vuitton:2007, Vuitton:2008, Cui:2009}. A list of molecules detected in Titan's atmosphere by Voyager, Cassini-Huygens, and ground-based telescopes is found in Table \ref{table:comp}. In particular, analyses of INMS data have discovered previously undetected nitrogen bearing molecules including HC$_{5}$N, NH$_{3}$, C$_{2}$H$_{3}$CN, etc.\ indicating that nitrogen plays a more prominent role in Titan's atmospheric chemistry than previously thought \citep{Vuitton:2007}. It is now clear that our pre-Cassini understanding of Titan atmospheric chemistry underestimated the chemical complexity of Titan's upper atmosphere. However, the limited mass range of INMS (1-100 Da/q)  does not allow for characterization of the larger molecules that are certainly involved in aerosol formation. 

The vertical profiles of molecular abundances have also been investigated in the stratosphere by CIRS and the mixing ratios of most species increase with altitude, which is a characteristic of molecules produced by photochemistry at higher altitudes that then condense in the lower stratosphere or at the tropopause (near 40 km where the temperature reaches a minimum of 70 K \citep{Fulchignoni:2005}) \citep{Vinatier:2007, Vinatier:2010}. This trend generally continues into the mesosphere and thermosphere as confirmed by VIMS, UVIS, and INMS measurements (see e.g., \citet{Adriani:2011, Koskinen:2011, Vuitton:2007, Cui:2009, Magee:2009}). Representative measurements from some of these instruments are shown in Figure \ref{fig:atmdata}. Latitudinal and seasonal variations will be discussed in Section \ref{Sect:season}.

In general, the vertical profiles of the trace species are well understood. One exception is the abundance of H$_{2}$. The tropospheric value ($1\times10^{-3}$, \citet{Niemann:2010}) and thermospheric values ($3.72\times10^{-3}$ at 1025 km, \citet{Yelle:2008, Cui:2009}) are incompatible by a factor of 2 \citep{Strobel:2010}, implying a downward flux of H$_{2}$ that is comparable to the escape flux. Measurements from CIRS and IRIS \citep{Samuelson:1981, Courtin:1995, Courtin:2012} are consistent with the results from GCMS in the regions where they overlap \citep{Courtin:2012}. This result is surprising given the long chemical lifetime and enormous escape flux of H$_{2}$ \citep{Strobel:2010, Strobel:2012}. \citet{Courtin:2012} find a non-uniform distribution of H$_{2}$ in the troposphere with an enhancement at high northern latitudes; this variation was not expected and may provide information required to resolve the apparent contradiction between the tropospheric and thermospheric measurements. 

Perhaps one of the most surprising and significant discoveries from Cassini has been the discovery of very heavy ions in Titan's upper atmosphere. Although INMS can only measure masses up to 100 Da/q, CAPS is able to detect much heavier masses, albeit at very low resolution. CAPS/ELS (Electron spectrometer) has measured negative ions up to $\sim$10,000 Da/q \citep{Coates:2007, Coates:2009}! The higher mass negative ions are observed preferentially at low thermospheric altitudes ($\sim$950 km) and tend to increase in mass with depth, at high latitudes, and in the region of the terminator \citep{Coates:2009, Coates:2010}. The extremely low mass resolution of the spectra has precluded identification of all but a few of the negative ions, CN$^{-}$, C$_{3}$N$^{-}$, and C$_{5}$N$^{-}$ \citep{Vuitton:2009}. CAPS/IBS (Ion Beam Spectrometer) has detected positive ions with masses up to 350 Da/q, that appear to have regular spacing between 12 and 14 Da, which indicates they are likely organic \citep{Crary:2009}. INMS data below 100 Da/q has been compared to the CAPS/IBS measurements and the composition has been estimated to include mostly aromatic compounds including naphthalene (C$_{10}$H$_{8}^{+}$) and anthracene (C$_{14}$H$_{10}^{+}$) in addition to some heteroaromatics \citep{Waite:2007, Crary:2009}. \citet{Delitsky:2010} argue that based on thermodynamics and laboratory experiments, high temperature reactions produce fused ring polyaromatic hydrocarbons (PAHs) while low temperature reactions will result in the formation of polyphenyls and accordingly some of the peaks in the CAPS spectrum could be biphenyl (154 Da), terphenyl (230 Da), and quaterphenyl (306 Da). Fullerenes have also been suggested as candidates for the negative ions \citep{Sittler:2009}. An unidentified feature at 3.28 $\mu$m in mesospheric measurements from VIMS \citep{Dinelli:2013} has been attributed to a collection of PAHs with an average mass of 430 Da \citep{Lopezpuertas:2013}. However, unambiguous identifications in this mass range will require a higher resolution mass spectrometer on a future mission. These large ions represent the transition from small gas phase molecules to haze particles; an ion with a mass of 10,000 Da could be a particle with a radius of $\sim$1.5 nm \citep{Lavvas:2013}, and therefore understanding the haze will require more information about these ions.

\section{Complex organic chemistry}
\subsection{Energy sources}
Energetic particles and solar ultraviolet photons dissociate and ionize N$_{2}$ and CH$_{4}$ initiating chemical reactions that result in the irreversible destruction of CH$_{4}$ leading to the formation of H$_{2}$ and a wealth of other complex organic molecules including the hydrocarbons and nitriles described above. The subsequent aggregation and heterogeneous chemistry of these molecules produce the aerosols responsible for Titan's characteristic orange color and distinctive haze layers. A schematic of the processes occurring in Titan's atmosphere is shown in Figure \ref{fig:titanatm}. 

There are a number of sources of ionization and dissociation in Titan's atmosphere: solar EUV and UV photons, energetic photoelectrons (resulting from solar X-ray and EUV photons), electrons from the Saturnian magnetosphere, other energetic ions (protons, oxygen), metallic ions created by micrometeorite ablation, and galactic cosmic rays (GCRs). Constraints on the energy available to drive chemical reactions and dynamics are of paramount importance for understanding Titan's atmosphere; instruments carried by Cassini-Huygens have significantly improved our knowledge of the distribution and relative importance of different sources of ionization and dissociation. The electron density can be measured by the Permittivity, Waves, and Altimetry analyzer (PWA) (40 to 140 km, \citet{Hamelin:2007, Lopezmoreno:2008}), radio occultations (above 120 km, \citet{Kliore:2008}), Radio and Plasma Wave Science (RPWS) (above 950 km, \citet{Wahlund:2005}), and the Cassini Plasma Spectrometer (CAPS) (ionosphere, \citet{Coates:2007, Coates:2010}). Additionally, UVIS measurements of EUV and FUV emissions \citep{Ajello:2007, Ajello:2008} place constraints on the N$^{*}$ and N$^{+*}$ densities ($^{*}$ indicates excited states), which are dissociation fragments of N$_{2}$. These instruments find that electron densities peak near 1200 km on the day side (2-4$\times10^{3}$ cm$^{-3}$, compared to 4-10$\times10^{2}$ cm$^{-3}$ on the night side) \citep{Agren:2009}. Day side ionization is dominated by solar photons, whereas night side ionization is dominated by Saturnian magnetospheric electrons \citep{Agren:2009, Galand:2010}. Energetic protons and oxygen ions are also a source of ionization from 500 to 1000 km \citep{Cravens:2008} and micrometeorite ablation results in ion formation from 500 to 700 km \citep{Molinacuberos:2001, Kliore:2008}. A secondary peak in the electron distribution is observed at 59 km (0.5-1$\times10^{3}$ cm$^{-3}$) \citep{Hamelin:2007, Lopezmoreno:2008} that results from the deposition of GCRs, which is comparable in ionization to solar photons in the upper atmosphere \citep{Gronoff:2009}. Note that based on measurements from Cassini's magnetometer, Titan does not currently have an internal magnetic field \citep{Backes:2005}.

The dominant source of energy to drive chemistry on Titan is solar photons, which are responsible for about 90\% of N$_{2}$ destruction \citep{Lavvas:2011b}. N$_{2}$ is the dominant source of absorption for photons with energies greater than 12.4 eV (100 nm), CH$_{4}$ for photons between 8.56 (145 nm) and 12.4 eV (100 nm), and the less energetic photons penetrate deeper in the atmosphere where they are absorbed by species produced from chemistry like C$_{2}$H$_{2}$ and C$_{4}$H$_{2}$ \citep{Lavvas:2011b}. C$_{2}$H$_{2}$ is particularly important because it absorbs above 5.17 eV (240 nm), a region where there is a significant photon flux. C$_{2}$H$_{2}$ and C$_{4}$H$_{2}$ play a particularly interesting role in photodissociation of methane. They absorb photons at lower energies than CH$_{4}$ and dissociate to form C$_{2}$H or C$_{4}$H, respectively, which then react with methane to produce methyl radicals (CH$_{3}$) and C$_{2}$H$_{2}$ and C$_{4}$H$_{2}$ as shown below:

C$_{2}$H$_{2}$ + h$\nu \rightarrow$ C$_{2}$H + H

C$_{2}$H + CH$_{4} \rightarrow$ CH$_{3}$ + C$_{2}$H$_{2}$

In this way, C$_{2}$H$_{2}$ and C$_{4}$H$_{2}$ behave as catalysts for methane destruction using photons that methane does not absorb and destroying methane in the stratosphere, much deeper than direct photolysis of CH$_{4}$ (see e.g., \citet{Yung:1984}).

\subsection{Atmospheric Chemistry \label{sec:chem}}

Photochemical models of Titan's atmosphere prior to Voyager predicted the presence of organic molecules resulting from CH$_{4}$ photochemistry \citep{Strobel:1974, Atreya:1978, Allen:1980}. The post-Voyager model of \citet{Yung:1984} approximately reproduced many of the observed column densities from Voyager but numerous reaction rates and products had to be estimated due to lack of experimental data and altitude profiles were not yet available from observations for comparison. In the decades between Voyager and Cassini-Huygens, improvements in reaction rate networks and improved observational constraints from analysis of Voyager data and Earth-based observations led to continued improvements in photochemical models (see e.g., \citet{Toublanc:1995a, Lara:1996, Wong:2002}). In preparation for Cassini-Huygens, \citet{Wilson:2004} developed a model that included the ionosphere and therefore included some ion chemistry, resulting in improved fits to pre-Cassini measurements.

\begin{figure}
{\includegraphics[width=0.9\textwidth]{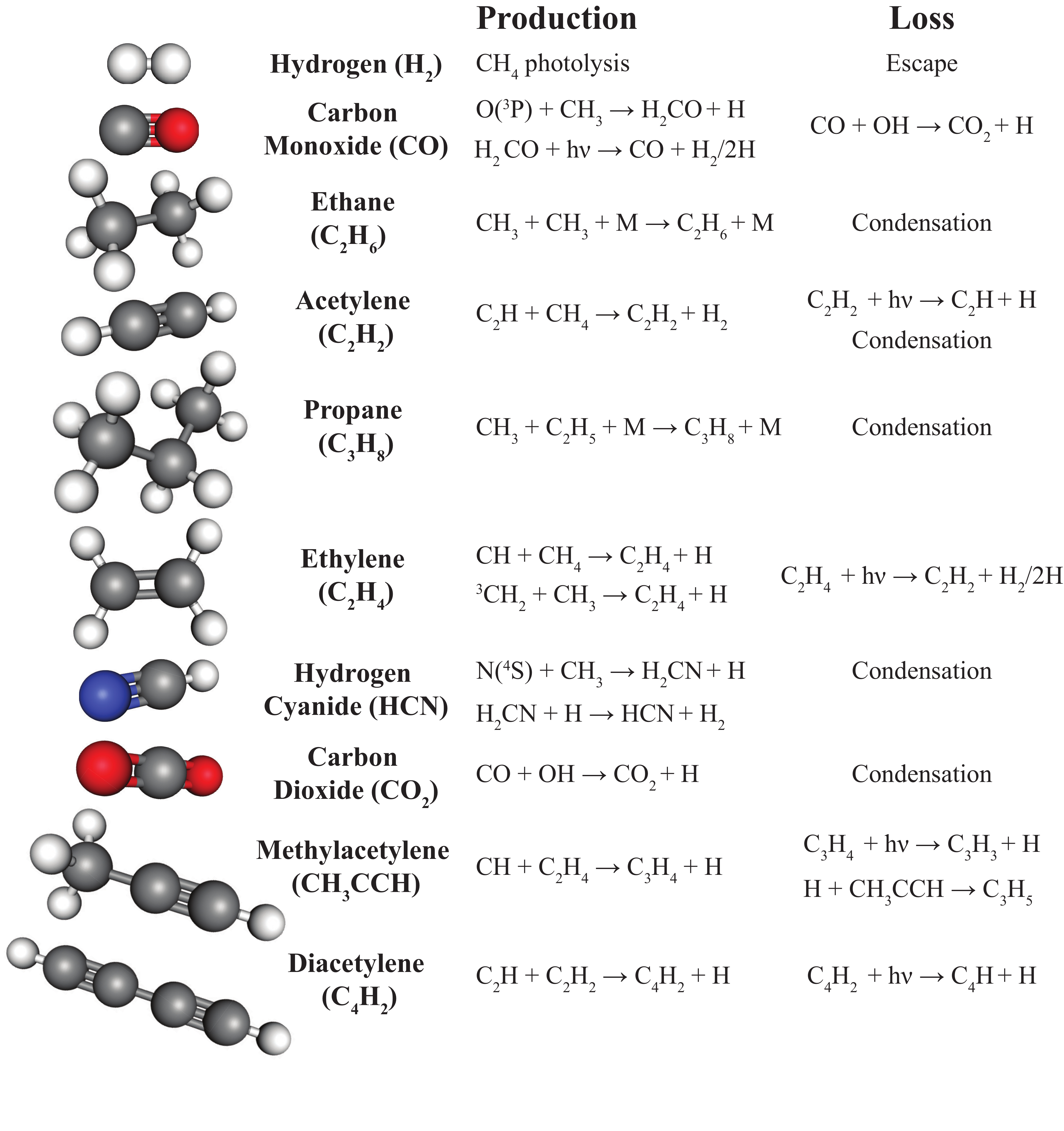}}
\caption{Shown here are the 10 most abundant photochemically generated molecules in Titan's equatorial stratosphere in approximately decreasing order of abundance with their major atmospheric production and loss pathways listed (see e.g., \citet{Vuitton:2014}).}
\label{fig:chem}
\end{figure}

One significant set of measurements from Cassini-Huygens are the composition measurements from mass spectrometry in the troposphere and thermosphere that have provided our first real constraints on the eddy diffusion profile in those regions of Titan's atmosphere \citep{Yelle:2008, Bell:2010, Cui:2012, Mandt:2012, Li:2014}, which is important because it determines the amount of time species spend in a given region of the atmosphere affecting their loss to reaction, photolysis, and condensation. For example, the relatively low eddy diffusion coefficient in the lower atmosphere allows large amounts of photochemically generated products to build up in the stratosphere, because transport to the deeper atmosphere where they are removed by condensation is slow; even so, as shown in Figure \ref{fig:chem} condensation is one of the main loss processes for molecules in Titan's atmosphere. The amount of material deposited on the surface varies depending on the model but in general models agree that liquids are more abundant than solids (by factors ranging from 1.5 to 14 depending on the model, \citet{Vuitton:2014}) and over the age of the solar system would produce a global layer of liquid a few hundred meters deep composed primarily of ethane with a layer of solids tens of meters deep. While these estimates are lower than predicted by post-Voyager models \citep{Lunine:1983}, they are still greater than what is observed on the surface as discussed in Section \ref{Sect:surface}.

Another significant improvement in our ability to model the chemistry in Titan's atmosphere comes from the characterization of aerosols discussed in Section \ref{sec:haze}; the distribution, size, and optical properties information obtained from Cassini-Huygens is an important input to any models of radiative transport in Titan's atmosphere and allows for more accurate treatment of photochemistry in models.

Photochemical models are now able to reproduce the observed abundances of most of the small (less than 100 Da) molecules in Titan's atmosphere (see e.g., \citet{Vuitton:2007, Horst:2008, Lavvas:2008a, Lavvas:2008b, Vuitton:2008, Yelle:2010, Krasnopolsky:2012, Vuitton:2012, Westlake:2012, Hebrard:2013, Dobrijevic:2014, Li:2015, Dobrijevic:2016}) but the formation of heavier molecules is not well understood. Figure \ref{fig:chem} shows a summary of the major production and loss pathways for 10 of the most abundant photochemically produced molecules in Titan's atmosphere (see extensive discussion in \citet{Vuitton:2014}).

Models generally agree that for the most part the hydrocarbon chemistry is driven by neutral reactions. However, to reproduce the INMS data, models need to take into account ion reactions as some species that are partially produced in the ionosphere, such as benzene (C$_{6}$H$_{6}$), may result from ion chemistry (C$_{6}$H$_{7}^{+}$ + e$^{-}$ $\rightarrow$ C$_{6}$H$_{6}$ + H) \citep{Vuitton:2008}. The nitrile chemistry begins with formation of HCN through neutral reactions and proceeds through the formation of CN from photolysis of HCN. CN can then react with acetylene or ethylene to produce cyanoacetylene or acrylonitrile respectively. Like benzene, formation of ammonia (NH$_{3}$) in the upper atmosphere seems to rely on the coupling of ion and neutral chemistry \citep{Yelle:2010}.

The oxygen containing molecules present in Titan's atmosphere (CO, CO$_{2}$, and H$_{2}$O) were discovered by Voyager and Earth-based telescopes \citep{Samuelson:1983, Lutz:1983, Coustenis:1998}, but their origin was uncertain. CO is a remarkably stable molecule, and its discovery in Titan's atmosphere \citep{Lutz:1983} led to investigations into whether the observed abundance is a primordial remnant, is supplied to the atmosphere from the interior or surface, or is delivered to the atmosphere from an external source.  The existence of remnant primordial CO would place useful constraints on models for the origin and evolution of Titan in the Saturnian nebula as discussed in Section \ref{sec:age}. Pre-Cassini models were unable to simultaneously reproduce the abundances of all three oxygen bearing species. CAPS measurements revealed a previously unknown source of oxygen in Titan's atmosphere, an external flux of O$^{+}$ ions \citep{Hartle:2006} flowing into the top of Titan's atmosphere, most likely from Enceladus. \citet{Horst:2008} demonstrated that this previously unknown source explains the presence of oxygen bearing species and the plumes of Enceladus are likely responsible for the fourth most abundant molecule in Titan's atmosphere (CO), revealing a unique connection between two very different moons in the Saturnian system. Subsequent measurements of H$_{2}$O \citep{Cottini:2012, Moreno:2012, Rengel:2014} have shown some disagreement with their model and further refinement may be necessary \citep{Krasnopolsky:2012, Moreno:2012, Dobrijevic:2014, Lara:2014}. The oxygen bearing species are unique in Titan's atmosphere because they include photochemically produced molecules that span atmospheric lifetimes of $\sim$10 yrs (H$_{2}$O) to $\sim$1 Gyr (CO) \citep{Horst:2008}; as such they provide an important opportunity to test photochemical models of Titan's atmosphere.

The O$^{+}$ ions observed by CAPS are deposited near 1100 km altitude, in the same region as the very heavy ions observed by CAPS. The addition of an oxygen source to complex organic compounds leads to the exciting possibility that molecules of prebiotic interest may form. Experiments indicate that amino acids and nucleobases may form under Titan atmospheric conditions \citep{Horst:2012}, but the instruments carried by Cassini-Huygens do not have a large enough mass range or sufficiently high resolution to detect such molecules.

The major positive ions in Titan's atmosphere are produced mostly through ion-molecule reactions that begin with CH$_{3}^{+}$ that form from direct ionization or charge transfer with N$_{2}^{+}$. HCNH$^{+}$ forms from N$^{+}$ + CH$_{4}$ or from proton transfer to HCN. Indeed a number of positive ion species form via proton transfer to abundant neutral species (see e.g., \citet{Keller:1998, Vuitton:2007, Carrasco:2008, Vuitton:2008}).

The presence of negative ions in Titan's ionosphere was not expected. Models of the troposphere looked at ion formation by electron attachment to radicals, which requires a three body collision and is therefore unlikely at the very low pressures in the ionosphere (see e.g., \citet{Capone:1976, Borucki:1987, Molina:2000}). However, as discussed above, negative ions with m/z up to 10,000 Da/q have been measured by CAPS. The model of \citet{Vuitton:2009} finds that CN$^{-}$ forms from dissociative electron attachment on HCN and HC$_{3}$N and C$_{3}$N$^{-}$ forms from proton transfer of HC$_{3}$N to CN$^{-}$. The negative ion chemistry on Titan has numerous sources of uncertainty: identifications from the CAPS measurements are difficult due to the low resolution of the instrument and lack of experimental measurements and/or theoretical calculations for negative ion production and destruction \citep{Vuitton:2014}.

Photochemical models of Titan's atmosphere generally only include species containing CHON atoms as those are the only ones thus far detected. Upper limits on PH$_{3}$ and H$_{2}$S were obtained from Cassini CIRS \citep{Nixon:2013} and additional efforts using ALMA to search for sulfur-bearing species are ongoing. Photochemical models indicate that better constraints on sulfur-bearing species may provide insight regarding external sources of material to Titan's atmosphere such as Enceladus, micrometeorites, and comets\citep{Hickson:2014}. Phosphorous and sulfur, in addition to CHON, are important from an astrobiological perspective and to improve our understanding of Titan's formation and evolution.

Improvements in understanding Titan's atmospheric chemistry require additional experiments and theoretical calculations. Models and experiments show that heterogeneous chemistry on the surfaces of aerosol particles may affect the gas phase composition (through adsorption, reaction with liquid/solid phase, and/or release of new gas phase species) \citep{Bakes:2003, Lebonnois:2003, Lebonnois:2005, Lavvas:2008b, Krasnopolsky:2009} but current models tend to include few, if any, of these processes. Modeling ionospheric chemistry is particularly challenging because it is difficult for experiments to access the conditions in Titan's upper atmosphere and therefore models lack numerous necessary measurements. The lack of measurements often leads people to ignore entire classes of reactions, such as dissociative electron attachment \citep{Vuitton:2009} and radiative association reactions \citep{Vuitton:2012} that turn out to be important when later included. Efforts to understand the largest sources of uncertainty in models will help focus limited resources on reactions that have the greatest impact (see e.g., \citet{Hebrard:2005, Hebrard:2013, Dobrijevic:2014}). Given the wealth of composition information from Cassini-Huygens, 3-D models will play an important role in improving our understanding of many of the seasonal changes discussed in Section \ref{Sect:season}.

\subsection{Photochemical haze \label{sec:haze}}

\subsubsection{Observations}
A brief note about nomenclature- the word aerosol refers to any particle suspended in a gas and is therefore the most general term to use to describe particles in Titan's atmosphere. Here ``cloud'' will be use to refer to a collection of particles that form by condensation of species in Titan's atmosphere (and may be solid or liquid phase). Haze will be used to refer to the photochemically generated involatile particles in Titan's atmosphere. It is often the case that ``tholin'' is used interchangeably with haze in the literature. However, for the sake of clarity, tholin is used here only to refer to laboratory generated analogues to Titan's photochemical haze (see \citet{Cable:2012}). 

The presence of an aerosol absorber in Titan's atmosphere was strongly suspected prior to the arrival of Pioneer 11 and the Voyager spacecraft based on ground-based observations that found a low geometric albedo in the ultraviolet \citep{Caldwell:1975}, a temperature inversion inferred from IR spectra \citep{Gillett:1975}, significant scattering in the atmosphere based on the morphology of the methane bands \citep{Trafton:1975}, and the inability to reproduce the optically thick atmosphere in the visible seen through polarization measurements using only gaseous absorbers \citep{Veverka:1973, Zellner:1973}. Images and polarization measurements from Pioneer 11 and Voyager 1 and 2 further constrained the properties of the aerosols in Titan's atmosphere and in combination with models suggested that Titan's aerosols are composed of organic material \citep{Rages:1980} and are fractal aggregates because two conflicting size dimensions (radii less than 0.1 $\mu$m and greater than 0.2 $\mu$m) were required to fit the data \citep{Rages:1983b, Tomasko:1980, West:1983, Hunten:1984, West:1991, West:1991b}. Images taken by Voyager 1 and 2 also revealed a North-South asymmetry in the haze with a $\sim$25\% decrease in brightness at blue wavelengths observed in the northern hemisphere \citep{Smith:1981, Smith:1982}. Vertical structure in the haze profile was also discovered in these images, including a detached haze layer between 340 and 360 km \citep{Rages:1983}.

Prior to the arrival of Cassini-Huygens, it was generally assumed that haze formed where it was observed, in the stratosphere \citep{Sagan:1984, McKay:2001, Lebonnois:2002, Wilson:2003}; instead, we now know that particles are present from Titan's surface to the ionosphere. Photochemically generated haze is the dominant source of opacity short of 5 $\mu$m but is optically thin in the IR and therefore plays a key role in the radiative balance and dynamics of Titan's atmosphere, acting to cool the surface while heating the stratosphere. The particles also affect and are affected by gas phase chemistry and can serve as condensation nuclei for cloud formation. Particles play a key role in atmospheric processes from the ionosphere to the surface, as discussed below.

As shown in Figure \ref{fig:haze}, stellar occultation measurements by UVIS revealed particles up to 1000 km \citep{Liang:2007, Koskinen:2011} and allowed retrieval of aerosol extinction coefficients up to 850 km \citep{Koskinen:2011}. Near-IR stellar and solar occultations observed by Cassini VIMS show evidence of fractal aggregates at altitudes above where DISR obtained measurements, a feature at 3.4 $\mu$m attributed to hydrocarbons that is not seen in laboratory analogues (discussed below), and a lack of NH and CN features at 3 and 4.6 $\mu$m, respectively, indicating that Titan's aerosol particles may have less nitrogen than their laboratory analogues \citep{Bellucci:2009, Kim:2011, Sim:2013, Kim:2013}. They identify CH$_{4}$, C$_{2}$H$_{6}$, CH$_{3}$CN, C$_{5}$H$_{12}$, C$_{6}$H$_{12}$, and C$_{6}$H$_{14}$ as possible constituents based on the VIMS measurements from $\sim$60-500 km \citep{Kim:2011, Sim:2013, Kim:2013}. Analysis of VIMS solar occultation measurements from 250-700 km show composition variations that may indicate aerosol aging and/or condensation onto particles as the particles become more aliphatic in character with decreasing altitude \citep{Courtin:2015}.

Below 300 km the haze is ubiquitous; though the distribution of particles varies as a function of altitude and latitude it does not vary spectrally in the far- and mid-IR indicating that chemical evolution measurable by CIRS is not occurring at a rate fast enough to compete with dynamics below 300 km (see e.g., \citet{Anderson:2011, Vinatier:2010b, Vinatier:2012}). The rate of dynamical mixing is not sufficient to homogenize the particle distribution though so it seems likely that most of the significant aerosol chemistry is occurring above 300 km. This is generally consistent with our understanding of chemical processes in Titan's atmosphere although it is potentially surprising given that gas phase chemistry still occurs at these altitudes. Measurements from CIRS and VIMS find that the signature of C-H and C-C bonds dominate the spectrum \citep{Rannou:2010,Vinatier:2012}. 

DISR provided the first ground truth on Titan's aerosol properties by measuring the absorption and scattering properties of the aerosols as a function of altitude. Many layers of condensates were predicted to form just above the temperature minimum at $\sim$ 44 km (see e.g., \citep{Sagan:1984}), however no distinct layers were observed with DISR in that region \citep{Tomasko:2005}. Their measurements are fit well by aerosol particles that are fractal aggregates composed of 4000 monomers with radii of 0.04 microns \citep{Tomasko:2008, Tomasko:2009}. DISR did not observe a significant vertical variation in the monomer size over their measurement range (surface to 140 km) \citep{Tomasko:2008}. Below 80 km, the observed change in the wavelength dependence of the optical depth and the single scattering albedo is attributed to the addition of condensate \citep{Tomasko:2009, Tomasko:2016}. There is also evidence of an increase in the volume extinction coefficient below 55 km \citep{Doose:2016}. DISR measurements below 80 km indicate that the aerosols are a combination of photochemically generated hazes and condensed organics \citep{Tomasko:2008}. In the visible, the particles' optical constants are roughly consistent with \citet{Khare:1984} but in the infrared Titan's particles are more absorbing than predicted from lab measurements \citep{Tomasko:2008}, which is also seen in CIRS measurements \citep{Vinatier:2012}. The haze model used to fit the DISR measurements can be used to predict the spectra seen by VIMS. At the Huygens landing site, the spectrum is a good fit to the VIMS data \citep{Tomasko:2009} and small changes in the ground albedo and haze thickness/absorption above 80 km provide good fits to the VIMS observations between -60$^{\circ}$ and 50$^{\circ}$ latitude \citep{Penteado:2010}.

Unlike on Earth, galactic cosmic rays (GCRs) do not penetrate all the way to Titan's surface. Theoretical models predicted that their deposition in the atmosphere would result in the creation of an ionosphere in the lower stratosphere \citep{Borucki:1987, Molina:1999, Molina:1999b, Borucki:2008}. Measurements from the mutual impedance (MI) and relaxation probe (RP) (part of the PWA) detected this ionosphere in the range of 50 to 80 km with a peak near 65 km (2$\times10^{9}$ m$^{-3}$ positive ions and 4.5$\times10^{8}$ m$^{-3}$ electrons) \citep{Fulchignoni:2005, Grard:2006, Hamelin:2007, Lopezmoreno:2008}. The presence of more positive than negative ions is consistent with the presence of negatively charged aerosols; the models of \citet{Lavvas:2010} and \citet{Larson:2014} find that the average particle charge density in the lower atmosphere should be 15 e$^{-}$/$\mu$m or 7.5 e$^{-}$/$\mu$m, respectively.

Huygens carried the Aerosol Collector Pyrolyzer (ACP), which collected a sample in the stratosphere and one in the troposphere, that were pyrolyzed, releasing gases to be analyzed by GCMS. GCMS detected HCN and NH$_{3}$, which may indicate that Titan's aerosol particles contain significant amounts of nitrogen \citep{Israel:2005, Israel:2006}, in apparent contradiction to the measurements from VIMS and CIRS discussed above. Note that others have argued for a different interpretation of the ACP measurements \citep{Biemann:2006}.

\subsubsection{Particle formation}

Given the discovery of very heavy ions and particles in the thermosphere, we now know particle production begins there. The haze production rate was constrained prior to the arrival of the Voyager spacecraft; \citet{Podolak:1979b} calculated an aerosol production rate of $3.5\times10^{-14}$ g cm$^{-2}$ s$^{-1}$ based on the observed optical properties of Titan's haze. More recent calculations also find values on the order of 10$^{-14}$ g cm$^{-2}$ s$^{-1}$ \citep{Mckay:1989, McKay:2001, Rannou:2003, Lavvas:2009, Lavvas:2011, Larson:2014}. \citet{Wahlund:2009} found an ionospheric mass flux of 3.2$\times10^{-15}$ g cm$^{-2}$ s$^{-1}$, demonstrating that a significant fraction of the mass of the haze particles is generated very high in the atmosphere. 

Chemistry involving aromatic molecules, benzene in particular, has been suggested as one possible pathway for aerosol formation based on chemical models (see e.g.,\ \citet{Lebonnois:2002, Wilson:2003, Lavvas:2008b, Delitsky:2010}) and laboratory experiments \citep{Trainer:2013, Sciamma:2014, Sebree:2014, Yoon:2014}. Models find that benzene production peaks in the ionosphere and has a higher relative abundance at higher altitudes \citep{Vuitton:2008}. Indeed most, if not all, of the proposed precursors (HCN, C$_{2}$H$_{2}$, C$_{2}$H$_{4}$, etc.) (see e.g.,\ \citet{Lebonnois:2002, Wilson:2003, Lavvas:2008b, Krasnopolsky:2009}) have larger mixing ratios in the ionosphere, where they are produced, than in the stratosphere. Titan's upper atmosphere is a much more energetic environment than the stratosphere, which may allow for greater degrees of nitrogen incorporation, important for prebiotic molecule formation, in aerosol particles. The specific chemical pathways leading to particle formation are still poorly understood; for example, models that use only neutral chemistry (see e.g., \citet{Lebonnois:2002, Wilson:2003}) produce particles too deep in the atmosphere.

Ionospheric chemistry results in the formation of small spherical particles. Using a starting point of 0.35 nm radius particles, \citet{Lavvas:2013} modeled their growth through chemistry and coagulation, as shown in Figure \ref{fig:haze}. They demonstrate that the particles are preferentially negatively charged and grow rapidly by addition of positive ions. They maintain their negative charge through electron photoemission and continue to grow. Their model explains a number of features from the CAPS observations: the striking difference in observed mass ranges for positive and negative ions, the high mass cutoff for negative ions, and the observed altitude dependent behavior. It is also consistent with the smaller than expected electron densities measured by the Langmuir Probe \citep{Agren:2012}. They conclude that Titan's lower ionosphere behaves as a dusty plasma and the contribution of particles must be included to fully understand Titan's ionosphere.

Deeper in the atmosphere, the aerosols form and grow by nucleation and deposition. As the atmospheric density increases and the sedimentation rate decreases, the aerosols grow by aggregation in the region of 650-500 km. Between 500 km and 400 km surface chemistry, which affects both composition and morphology, dominates as the aerosols react directly with gas phase radicals such as HCCN \citep{Lavvas:2011}. Photochemical models do not generally include heterogeneous chemistry on the surfaces of the aerosols due to lack of necessary laboratory measurements, however \citet{Sekine:2008, Sekine:2008b} showed that the atomic hydrogen abundance in Titan's atmosphere can be significantly affected by addition or recombination processes on the surfaces of the aerosols. Additionally, the inclusion of heterogeneous chemistry can act to reduce the abundances of saturated hydrocarbons, decrease HCN, and increase the abundance of phenyl radicals \citep{Sekine:2008, Sekine:2008b}.

Once the particles reach the stratosphere and troposphere, they are subject to dynamical processes and condensation described elsewhere in this work. Indeed, \citet{Rannou:1997} noted that microphysics alone could not explain the observed properties of Titan's haze and therefore dynamics must also play a role. Given that the particle growth processes are altitude dependent, it follows that the composition of the particles likely varies with altitude at least until they reach the stratosphere. This set of processes is similar to the qualitative description of \citet{Sagan:1984}, although in their scenario haze formation occurred at a much lower altitude and results in a maximum heavy atom number of 9 or 10 ($\sim$130 Da) in the haze. 

The resulting particles are fractal aggregates with a fractal dimension of 2, intermediate between linear aggregates (1) and spherical aggregates (3) (see e.g.,\ \citet{Cabane:1993, Rannou:1997, Lavvas:2009}). The morphology is important since aggregates coagulate faster and fall slower because they have a larger cross sectional area for a given mass \citep{Cabane:1993, Larson:2014} and fractals absorb more in the UV to mid visible than spherical particles of the same mass and have different optical properties because light interacts with multiple dimensions \citep{Lavvas:2010, Larson:2014}. Microphysical models are able to reproduce the properties of the haze inferred from DISR including number of monomers, particle size, and density. However, below 80 km the aerosol characteristics require the addition of mass with different optical properties. This is probably indicative of condensation \citep{Lavvas:2010, Lavvas:2011c} as discussed in Section \ref{Sect:clouds}. \citet{Larson:2014} note that with their model it takes particles 395 years to get from the mesosphere to the surface.

\subsubsection{Particle composition}

Based on assumptions regarding the molecular precursors to haze, photochemical models can estimate the composition of the haze particles as a function of altitude. Frequently used precursors include aliphatic copolymers, nitriles, aromatics, and polyynes. Even if the same precursors are used, models differ on the relative importance of the classes of molecules involved in haze formation. \citet{Wilson:2003} found that haze formation is dominated by aromatics peaking in production at 220 km, with a smaller production peak for nitriles at $\sim$900 km. Other models have found that the total production of aerosols is dominated by aliphatic copolymers in the upper atmosphere and nitriles in the lower atmosphere (a result of energy deposited by GCRs) \citep{Lavvas:2008a, Lavvas:2008b}. The C/N ratio of the haze is then estimated and ranges from 1.3-2 \citep{Lara:1999, Lavvas:2008b} to 16.7 \citep{Krasnopolsky:2009} and the estimated C/H ratio ranges from 0.83 \citep{Krasnopolsky:2009} to 1.3 \citep{Lavvas:2008b}. The wide range in values can be attributed to assumptions made about the chemical pathways involved in haze formation. \citet{Lebonnois:2003} calculated the C/N ratio of the aerosols based on 3 previously suggested pathways. Polymers of C$_{2}$H$_{2}$ and HC$_{3}$N \citep{Clarke:1997} result in C/N of 10-20 depending on the altitude, PAHs with HCN and HC$_{3}$N resulted in a C/N of 8-20, and polymers of HCN and other nitriles result in a C/N of 1.5. Other laboratory haze simulation experiments, discussed below, tend to produce particles with C/N values ranging from 1 to 3.5 (see e.g.,\ \citet{Sarker:2003, Imanaka:2004, Imanaka:2010, Sciamma:2010, Trainer:2012}), although it increases with inclusion of trace aromatics \citep{Trainer:2013, Yoon:2014}. In all cases, these models and experiments show that the haze may be an important sink for both hydrogen and nitrogen in the atmosphere, which is consistent with the results from Huygens ACP \citep{Israel:2005, Israel:2006} but inconsistent with the lack of nitrogen in the Cassini CIRS and VIMS measurements mentioned above (see e.g.,\ \citet{Rannou:2010,Vinatier:2012}).

Laboratory experiments, with their ability to control conditions and access to state-of-the-art analytical techniques, play an important role in understanding Titan's haze. Titan haze analogues are often called tholins, a term first coined by \citet{Sagan:1979}. They wrote,``The product, synthesized by ultraviolet (UV) light or spark discharge, is a brown, sometimes sticky, residue, which has been called, because of its resistance to conventional analytical chemistry, `intractable polymer.'...It is clearly not a polymer-- a repetition of the same monomeric unit-- and some other term is needed...We propose, as a model-free descriptive term `tholins'..., although we were tempted by the phrase `star-tar'." The word tholin does not refer to a unique material, rather a class of materials that are produced under a range of conditions but share the general characteristics of starting with a cosmically abundant gas or ice and exposing it to an energy source that results in the formation of a refractory solid material. For a broad overview of tholin work, see \citet{Cable:2012}.

Although the optical properties of a variety of tholins have been measured \citep{Podolak:1979, Khare:1984, Bar-Nun:1988, Coll:2001, Ramirez:2002, Tran:2003, Bernard:2006, Quirico:2008, Vuitton:2009b, Hasenkopf:2010, Gautier:2012, Imanaka:2012, Sciamma:2012, Mahjoub:2012, Hadamcik:2013}, no laboratory analogue completely reproduces Titan's haze optical properties over the measured wavelength regions \citep{Tomasko:2008, Bellucci:2009, Lavvas:2010, Rannou:2010, Anderson:2011, Kim:2011, Vinatier:2012, Hirtzig:2013, Kim:2013, Sim:2013, Courtin:2016}. \citet{Brasse:2015} review the current state of optical constants for tholins. It is important to be aware that the optical constants of \citet{Khare:1984} are used most frequently because they are the only in depth study of optical constants for planetary haze particles that span a wide spectral range. The \citet{Khare:1984} optical constants are particularly bad fits to Cassini measurements at 3, 3.4, 13-17, and 72 $\mu$m. Although some Titan tholins do have features at 3.4 $\mu$m, the retrieved value from Cassini VIMS is significantly higher than what is measured in the lab \citep{Rannou:2010}. The observations at 3.4 $\mu$m are complicated by a significant contribution from C$_{2}$H$_{6}$ gas at that wavelength \citep{Maltagliati:2015}. The regions of spectral agreement indicate that there are conjugated structures (often the result of alternating single and double bonds) in Titan's haze particles \citep{West:2014}. \citet{Sebree:2014} recently showed that the addition of aromatic molecules during tholin production results in a better fit to the feature seen at 140 cm$^{-1}$ in CIRS, indicating aromatic molecules are likely present in the haze.

Experiments have shown that photoionization of N$_{2}$ is important for the nitrogen chemistry and the formation of heavy organics such as benzene and toluene because the yields of those molecules increase at shorter wavelengths \citep{Imanaka:2007, Thissen:2009}. It has also been noted based on comparisons of gas and solid products that nitrogen appears to be very efficiently incorporated into the solid phase \citep{Imanaka:2010}. Indeed a number of experiments have demonstrated that the presence of N$_{2}$ increases the particle production rate in laboratory simulations \citep{Imanaka:2010, Trainer:2012}. Further, work by \citet{Horst:2014} shows that the presence of CO also increases particle production rates. The combination of N$_{2}$, CH$_{4}$, and CO in Titan's atmosphere make it an ideal place for photochemical generation of particles. 
 
The wealth of \emph{in situ} and remote sensing observations obtained by Cassini-Huygens has provided new constraints on Titan's aerosols and indicates that their formation initiates much higher in the atmosphere than previously believed. The aerosols therefore carry the signature of upper atmosphere processes, though they undergo additional chemical and physical processing as they settle to the surface. During this time the initially spherical particles eventually form fractal aggregates, which can serve as condensation nuclei for molecules that condense in the lower atmosphere such as HCN, C$_{2}$H$_{6}$, and CH$_{4}$. Since the aerosols are produced globally, they are eventually deposited onto a wide variety of surface environments, possibly allowing for further alteration under different conditions including processing in the lakes and seas or in cryovolcanic flows or impact generated melt pools. Experiments have shown that tholins produced from mixtures of N$_{2}$ and CH$_{4}$ can be hydrolyzed to form molecules of prebiotic interest including amino acids (the building blocks of proteins) \citep{Neish:2010, Poch:2012}.

\section{Dynamics \label{sec:dynamics}}

\subsection{Winds \label{Sect:winds}}

The winds in Titan's atmosphere result primarily from solar forcing, which varies seasonally. There are a number of different ways to constrain wind speeds on Titan. Direct measurements of the wind speeds come from a few methods, each with limitations: measurements from the Doppler Wind Experiment (DWE) carried by the Huygens probe \citep{Bird:2005, Folkner:2006}, measuring doppler shifts in the emission lines of atmospheric constituents like ethane (see e.g., \citet{Kostiuk:2001, Kostiuk:2005, Luz:2005, Moreno:2005, Kostiuk:2006, Luz:2006, Kostiuk:2010}), and cloud tracking \citep{Bouchez:2005, Porco:2005, Turtle:2011}. The DWE provided measurements at one place and time from 145 km to the surface. The doppler shift technique has the advantage of being possible from Earth, but provides limited spatial resolution. Cloud tracking is often used to measure wind speeds in planetary atmospheres (see e.g., \citet{Vasavada:2006}) but the general paucity of clouds on Titan means that cloud tracking provides temporally and spatially limited wind speed measurements. Indirect techniques have also been used to constrain the winds on Titan including use of the thermal wind equation to calculate wind speeds from temperature measurements (see e.g., \citet{Flasar:2005, Achterberg:2008}) and stellar occultations \citep{Sicardy:1990, Hubbard:1993, Bouchez:2004, Sicardy:2006}. Wind speeds can also be constrained through evidence of aeolian processes such as dunes and wind-driven waves on the lakes and seas as discussed in Section \ref{Sect:surface}. Of all these techniques, only use of the thermal wind equation on CIRS temperature measurements provides us with significant amounts of temporal and spatial information including seasonal changes (see e.g., \citep{Achterberg:2011}). 

\begin{figure}
{\includegraphics[width=0.5\textwidth]{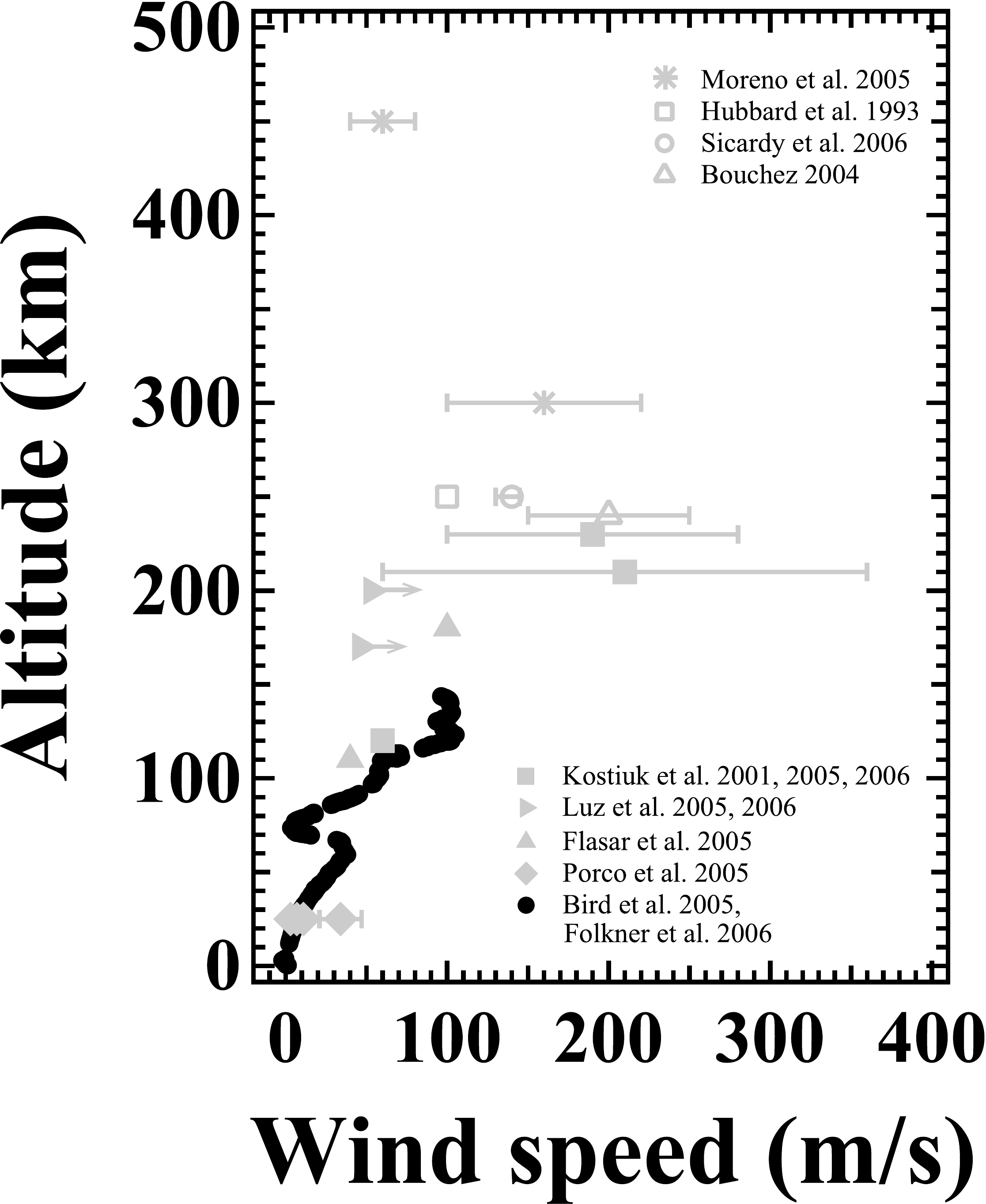}}
\caption{Measurements from various techniques reveal that wind speeds are relatively low in the troposphere but in the stratosphere there are superrotating winds. Black circles are the measurements from the Doppler Wind Experiment (DWE) \citep{Bird:2005, Folkner:2006}. Cloud tracking measurements from \citep{Porco:2005} are shown as filled diamonds; the altitude for these clouds is not well constrained and shown here is a presumed upper limit. The measurements made from doppler shifts in the emission lines of atmospheric constituents are shown as filled squares \citep{Kostiuk:2001, Kostiuk:2005, Kostiuk:2006}, filled right pointing triangles (arrows indicate these measurements are lower limits) \citep{Luz:2005, Luz:2006}, and asterisks \citep{Moreno:2005}. \citet{Hubbard:1993} (open square), \citet{Sicardy:2006} (open circle), and \citet{Bouchez:2004} (open triangle) used stellar occultations to estimate the stratospheric wind speeds. Wind speeds calculated from the thermal wind equation using Cassini CIRS temperature measurements near the Huygens' landing site are shown as triangles \citep{Flasar:2005}. Note that although there is some spatial and temporal overlap, this plot does not necessarily represent a snapshot of Titan's atmosphere at one place and time.}
\label{fig:wind}
\end{figure}

The wind speed profile obtained by the DWE is shown in Figure \ref{fig:wind}. In the troposphere, cloud tracking measurements, which are generally consistent with the DWE, find wind speeds primarily eastward ranging from 0.5 to 10 m/s \citep{Bouchez:2005, Turtle:2011}, although speeds up to 34 $\pm$ 13 m/s have been reported \citep{Porco:2005}. Higher in the atmosphere, Titan exhibits stratospheric superrotating winds (up to $\sim$200 m/s) \citep{Flasar:1981, Flasar:2005, Achterberg:2008, Sicardy:1990, Hubbard:1993, Kostiuk:2001, Bouchez:2004, Kostiuk:2005, Luz:2005, Kostiuk:2006, Sicardy:2006, Luz:2006}, which then decrease in the upper stratosphere and lower mesosphere \citep{Moreno:2005, Achterberg:2008}. Taken together, the zonal wind speeds are very low ($<$1 m/s) at the surface increasing to $\sim$40 m/s near 60 km where they begin decreasing until they reach a minimum ($\sim$5 m/s) around 75 km before increasing again to speeds up to 200 m/s near 200 km then decreasing to 60 m/s at higher altitudes (near 450 km). \citet{Li:2012} referred to this minimum as the ``zonal wind collapse'' and although some atmospheric models find a minimum in the wind profile in this region of the atmosphere, they are unable to reproduce the magnitude and/or altitude of the feature \citep{Lebonnois:2012, Li:2012}.

\citet{Achterberg:2008} averaged CIRS measurements from 2004 to 2006 (Northern winter) and showed that the stratospheric jet was centered between 30$^{\circ}$N and 50$^{\circ}$N at a pressure of $\sim$0.1 mbar and argue that although pre-Cassini temporal coverage is sparse, it is consistent with the presence of a strong jet in the winter hemisphere, which forms sometime in the fall and dissipates after spring equinox. The length of the Cassini-Huygens mission and the spatial coverage of Cassini CIRS enables monitoring of Titan's stratospheric winds as the seasons change. \citet{Achterberg:2011} showed that from 2005 to 2009, the southern hemisphere zonal winds exhibited very little change in the stratosphere. In the northern hemisphere, the superrotating jet extended to higher altitudes.

The meridional wind speeds have been estimated based on images taken by DISR in the lower troposphere (up to $\sim$20 km) to be an average of $\sim$0.1 m/s northward above 1 km (peaking at 0.4 m/s) and southward below 1 km (peaking at 0.9 m/s) \citep{Karkoschka:2016}. These results are consistent with estimates of the very near surface wind speeds (north-south, 0.4 m/s) made through analysis of tracking the parachute \citep{Karkoschka:2007}, cooling of the probe on the surface \citep{Lorenz:2006b}, and conclusions about the motion of the Huygens probe as it landed made from a synthesis of the various data it acquired \citep{Schroder:2012}. The meridional winds are difficult to measure; however, meridional circulation transports both trace gas species and particles, which can used as tracers of the circulation as discussed below.

\subsection{General Circulation} 

Titan's slow rotation allows for global Hadley circulation, with ascending motion in the summer hemisphere and descending motion near the winter pole, to redistribute heat. As with Earth, Titan's large obliquity ($\sim$26.7$^{\circ}$) results in seasonal variations in solar forcing. Unlike Earth, where the strength of the Coriolis force limits the latitudinal extent of meridional circulation cells, Titan generally has 1 main cell (pole-to-pole), except near the equinoxes when there are two cells (equator-to-pole) as the circulation reverses (see e.g., \citet{Hourdin:1995, Mitchell:2009b, Newman:2011}), which led \citet{Mitchell:2006} to describe Titan as having an ``all tropics'' climate. The single Hadley cell efficiently redistributes heat resulting in relatively small equator to pole temperature contrasts \citep{Jennings:2009, Schinder:2011, Cottini:2012, Schinder:2012}. The surface temperature decreases asymmetrically from the equator to the poles with a decrease of 3 K toward the winter pole (60$^{\circ}$N) and 1 K toward the summer pole (60$^{\circ}$S) \citep{Cottini:2012}. In terms of atmospheric circulation, Titan is intermediate to Earth and Venus as its atmospheric pressure is higher and rotation rate slower than Earth but lower and faster than Venus; however like Earth, Titan has a significant obliquity.

Most GCMs of Titan's atmospheric dynamics predict the Hadley cell extends from the surface to the stratosphere. However, based on measurements of temperature and chemical tracers, \citet{Teanby:2012} argue that the Hadley circulation may extend up as far as 600 km from the surface, above the top boundary of the GCMs. Limitations on the placement of the top boundary of GCMs are frequently invoked as the explanation for mismatches between models and a variety of observations. The winds near the surface have been interpreted as direct evidence of the surface branch of the Hadley circulation, which is predicted to extend to an altitude of a few km (see e.g., \citet{Tokano:2007, Friedson:2009, Mitchell:2011, Schneider:2012, Charnay:2012}); a reversal seen near 1 km in the Huygens data \citep{Karkoschka:2007} may correspond to the upper boundary of the surface branch \citep{Lebonnois:2014}. In addition to the main Hadley circulation, a number of GCMs find that a small tropospheric cell forms in the summer hemisphere \citep{Newman:2011, Lebonnois:2012, Lora:2015b}. The latitudinal extent of the global Hadley cell depends on certain aspects of the model (see e.g., \citet{Mitchell:2009b, Lora:2011, Newman:2016}); \citet{Lora:2011} find that the hottest point in the summer hemisphere does not reach the pole and the upwelling branch does not either. The latitude profile of the zonal winds is key for understanding horizontal angular momentum transport processes, but as discussed in Section \ref{Sect:winds}, winds on Titan are often poorly constrained. Cloud tracking measurements tend to suggest higher wind speeds at the midlatitudes than at the poles, which potentially indicates that angular momentum is uniform with latitude and would be consistent with regions dominated by Hadley circulation \citep{Lebonnois:2014}. 

Observations that Titan's spin rate differs from synchronous \citep{Lorenz:2008d, Stiles:2008, Stiles:2010} have been explained by the exchange of angular momentum between Titan's massive atmosphere and its surface assuming the presence of a subsurface ocean \citep{Lorenz:2008d}. This exchange may lead to seasonal variations in Titan's rotation rate because the seasonal reversal of the meridional circulation produces a torque; the exact details of this interaction are still being explored (see e.g., \citet{Tokano:2005, Karatekin:2008, Lorenz:2008d, Mitchell:2009, VanHoolst:2009, Goldreich:2010, Tokano:2011, Richard:2014, Tokano:2014b}). 

In Titan's atmosphere the radiative time constant decreases rapidly with increasing altitude. In the troposphere, it is long compared to seasonal time scales (10 Titan years or longer, \citet{Strobel:2009}). In the stratosphere and mesosphere, the radiative time constant is much shorter than a Titan season. The winter polar lower stratosphere ($\sim$100-170 km) is an exception, as the radiative time constant there is comparable to the seasonal timescale \citep{Achterberg:2011}. The shorter radiative time constant means that temperatures adjust to seasonal forcing in the stratosphere as observed in CIRS data (see Section \ref{Sect:season}). Interestingly, \citet{Cottini:2012} found a diurnal temperature variation of $\sim$1.5 K near the equator and seasonal changes in the surface temperature have also been observed \citep{Jennings:2011} indicating that the thermal inertia of the surface must be low. It is important to note that Saturn's orbit is sufficiently eccentric (0.054) to result in seasonal asymmetries as the solar flux at Titan varies by $\sim$20\% throughout Saturn's orbit. Currently Titan is closest to the sun at southern summer solstice. The orbital asymmetries are dominated by precession of perihelion which has a timescale of $\sim$45,000 yr \citep{Roe:2012b}.

Another timescale of interest for understanding Titan's atmospheric circulation is the dynamical timescale. For global circulation, the meridional overturning timescale, the meridional scale divided by the mean meridional velocity, is the timescale of interest. Estimations of these timescales require GCMs. Using Titan WRF, \citet{Newman:2011} find that the shortest dynamical time constants (a few Titan days) exist in the upper stratosphere at low latitudes where the meridional winds are fastest and the longest (a few Titan years) in the upper troposphere and lower stratosphere at mid to high latitudes \citep{Lebonnois:2014}. These estimates are consistent with the less complex model of \citet{Flasar:1981} based on Voyager measurements. 

A comparison of radiative and dynamical timescales indicates that the tropospheric and lower stratospheric radiative timescales are longer or comparable to the dynamical timescales, which is consistent with the observed efficiency of heat redistribution resulting in very small equator to pole temperatures variations in Titan's atmosphere. Higher up in the stratosphere, the meridional dynamical timescales are longer and radiative processes should dominate, but the changes in heating also result in changes in zonal winds to maintain the thermal wind balance. As a result, seasonal changes in stratospheric temperatures and winds are expected and consistent with observations \citep{Lebonnois:2014}

In addition to heat, the meridional circulation transports angular momentum resulting in high latitude, high altitude jets. In the stratosphere, this allows the centrifugal force to equilibrate the latitudinal pressure gradient force (PGF) in place of the Coriolis force and the atmosphere is in the cyclostrophic regime \citep{Flasar:1981}. The troposphere is in the geostrophic regime (where pressure gradient force balances the Coriolis force) and the transition between the two regimes occurs somewhere mid-troposphere where the balance is between the PGF, Coriolis, and centrifugal forces \citep{Mitchell:2016}.

One interesting feature of Titan's atmospheric structure is that the axis of symmetry for winds and temperatures is tilted 4 degrees with respect to the pole - tilted towards the sun in the northern hemisphere and shifted 76$^{\circ}$W from the subsolar longitude \citep{Achterberg:2008b}. A similar tilt (3.8 degrees, 79$^{\circ}$W) is observed in the hemispheric brightness asymmetry seen in Cassini imaging data \citep{Roman:2009} and is required to fit the measured HCN distribution from Cassini CIRS \citep{Teanby:2010}. The HCN ice cloud at Titan's south pole and distribution of trace species such as HC$_{3}$N and C$_{4}$H$_{2}$ also exhibit a 4 degree offset from the south pole \citep{Jennings:2015}. \citet{Tokano:2010b} suggest that thermal tides may be responsible for the tilt but the explanation is not consistent with all of the observations at equinox when it predicts that the tilt decreases to 1 degree. Deeper investigation of the evolution of the tilt throughout the course of the Cassini mission should provide additional constraints on possible explanations \citep{Achterberg:2008b}.

\subsection{Superrotation}

As discussed above, one of the defining characteristics of Titan's atmospheric circulation is the presence of superrotating winds in the stratosphere including a strong jet near 0.1 mbar ($\sim$300 km) in the winter hemisphere. The first Titan GCM to successfully generate superrotation was \citet{Hourdin:1995} demonstrating that numerical angular momentum conservation is important for generation of superrotation in models. However, other Titan GCMs, especially early on, have difficulties reproducing this superrotation \citep{Tokano:1999, Richardson:2007, Friedson:2009, Larson:2014}  and although a number of different GCMs can now reproduce superrotation (though not always strong enough) \citep{Newman:2011, Lebonnois:2012, diaspinto:2014, Lora:2015b, Newman:2016}, we lack consensus as to how to maintain it \citep{Lora:2015b}. Note the problems with generating and sustaining superrotation are not unique to Titan. The Venus atmospheric modeling community has faced similar difficulties (see e.g., \citet{delgenio:1996, Zhu:2006, Lebonnois:2010}). Unlike Venus, Titan has a strong seasonal cycle, which complicates the atmospheric dynamics, and \citet{Mitchell:2014} showed that strong seasonal effects act to prevent generation of superrotating winds and needed to adjust the Newtonian cooling time to achieve superrotation.

The results of models that generate superrotating winds (see e.g., \citet{Hourdin:1995, delgenio:1996, Newman:2011, Lebonnois:2012}) seem to support the idea that the mean meridional circulation transports excess angular momentum upward and poleward and that barotropic waves generated by instabilities on the edges of the high latitude jets transport momentum towards the equator \citep{Lebonnois:2014}, which is known as the GRW mechanism \citep{Gierasch:1975, Rossow:1979}. For example, some models find that superrotation requires coupling between equatorial and high latitude waves but that generation and maintenance require equatorial Kelvin-like waves and Rossby-like waves, respectively \citep{diaspinto:2014, Wang:2014}. Though waves play a key role in the atmospheric dynamics, they are difficult to constrain; observations of clouds may play a key role in revealing the characteristics of waves in Titan's atmosphere \citep{Mitchell:2011}.

Superrotation and zonal jets depend on the mean meridional circulation, which is determined by rotation rate, planetary radius, and solar heating but it is also affected by radiative transport and is therefore also related to the gas, haze, and cloud distributions. Models indicate that coupling of gas composition and dynamics is far less important than coupling of haze and dynamics \citep{Lebonnois:2003, Hourdin:2004}, though it is important to remember that the particle and gas phases are connected in Titan's atmosphere through photochemistry and condensation processes. Coupling with haze increases the maximum speeds achieved by zonal winds as transport of particles increases the equator to pole temperature variations thus strengthening the meridional circulation (see e.g., \citep{Rannou:2002, Rannou:2004, Lora:2015b}). 

\section{Seasonal change \label{Sect:season}}

\begin{figure}
{\includegraphics[width=0.9\textwidth]{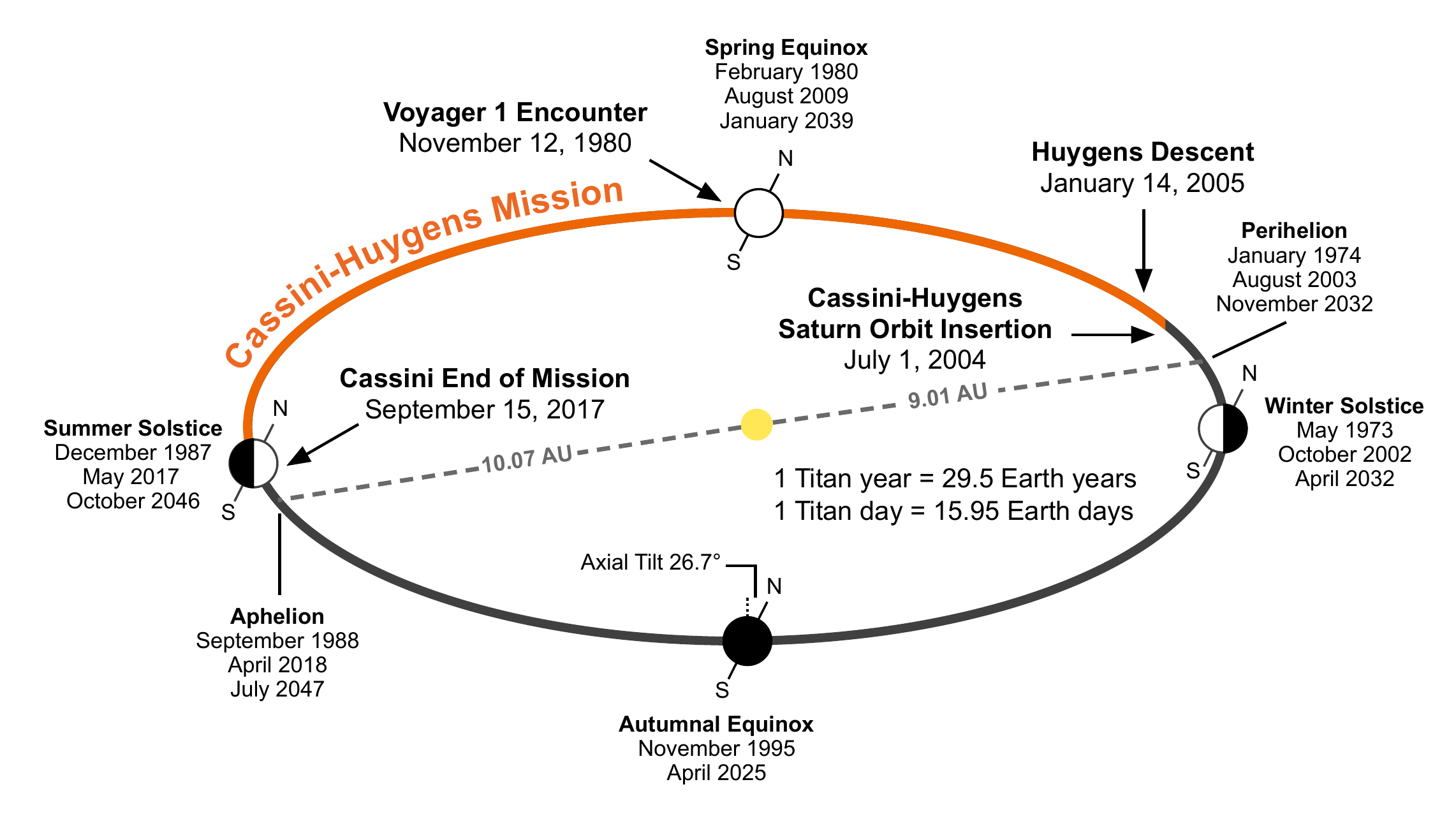}}
\caption{Shown here is schematic of Titan's orbit, which notes the length and timing of the Cassini-Huygens mission.}
\label{fig:seasons}
\end{figure}

Seasonal effects are important over a Titan year (29.5 Earth years) because of Saturn's axial tilt of almost 27$^{\circ}$. Due to the length of the Cassini mission, seasonal changes can be investigated (see Figure \ref{fig:seasons}). In addition, the seasonal overlap with Voyager allows for some investigation of inter-annual variability.

\subsection{Composition and temperature}

\begin{figure}
{\includegraphics[width=0.9\textwidth]{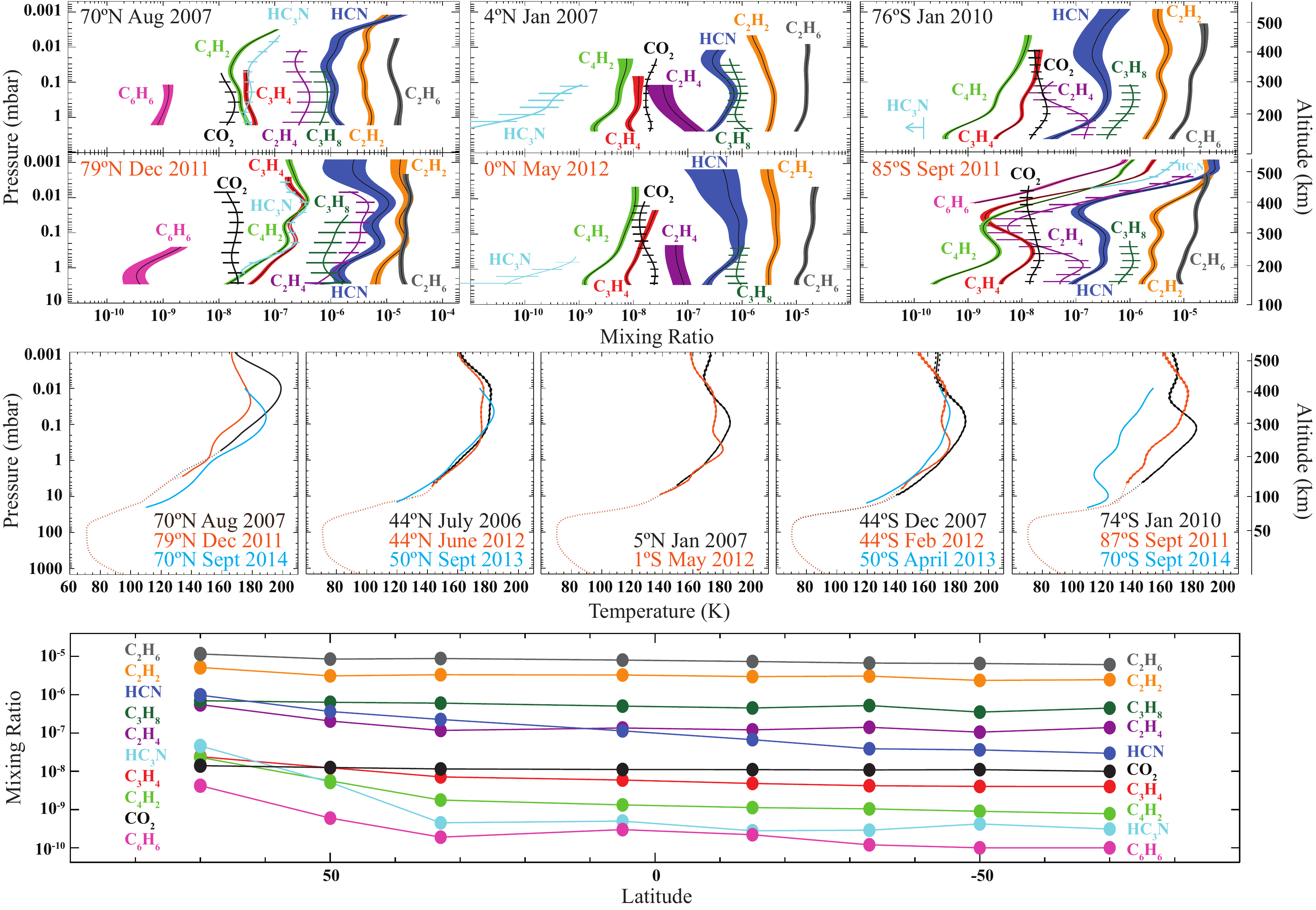}}
\caption{This compilation of Cassini CIRS measurements adapted from \citet{Coustenis:2010}, \citet{Vinatier:2015}, and \citet{Coustenis:2016} shows variations in stratospheric composition and temperature with latitude and season. The top panels compare limb composition measurements at northern, equatorial, and southern latitudes before and after the August 2009 equinox (with the exception of the south pole which shows that the significant changes on composition happen post-equinox) \citep{Vinatier:2015}. The middle panel shows the evolution of temperature with season and compares temperature profiles near 70$^{\circ}$N, 44$^{\circ}$N, the equator, 44$^{\circ}$S, and the south pole \citep{Vinatier:2015, Coustenis:2016}. The bottom panel shows the  CIRS nadir measurements (with contribution function peaks near $\sim$5-10 mbar) from 2004 to 2008 demonstrating the significant north (winter) polar enrichment of short lived species \citep{Coustenis:2010}. For the top and middle panels, the altitude scale is from the equatorial measurements; pressure is the relevant coordinate and applies to all panels. For the temperature profiles, the dashed line indicates regions of the atmosphere where CIRS limb measurements do not constrain the temperature (see \citet{Vinatier:2015}).}
\label{fig:temps}
\end{figure}

As shown in Figure \ref{fig:temps}, both Voyager IRIS and Cassini CIRS measurements show strong variations with latitude for stratospheric mixing ratios and in general hydrocarbons and nitriles are more abundant at the winter pole than at lower latitudes or in the summer hemisphere \citep{Coustenis:1995, Coustenis:2007, Vinatier:2007,Teanby:2008, Teanby:2008b, Teanby:2012, Vinatier:2015, Coustenis:2016}. This likely results from the winter polar vortex, which enhances abundances through subsidence of air from higher altitudes, where the molecules are produced and have larger mixing ratios, to deeper regions where they are protected from photolysis by the winter polar night (see e.g., \citet{Yung:1987, Lebonnois:2001, Hourdin:2004, Crespin:2008,Teanby:2008}). This trend was observed when Cassini first arrived during northern winter, but has now been observed at the south pole as it moves into winter \citep{Teanby:2012, Bampasidis:2012, Vinatier:2015, Coustenis:2016}. Equinox occurred on 11 August 2009 and evidence of seasonal changes in the form of increasing trace gas abundances and temperature changes at the south pole were first observed in June 2011 \citep{Teanby:2012}. The strength of the latitudinal variation depends on the chemical lifetime of the species (see e.g., \citet{Lavvas:2008a}); short lived species (e.g., C$_{2}$H$_{4}$, C$_{3}$H$_{4}$, C$_{4}$H$_{2}$, C$_{6}$H$_{6}$, HC$_{3}$N) tend to exhibit stronger variations than long lived species (e.g., C$_{2}$H$_{2}$, C$_{2}$H$_{6}$, C$_{3}$H$_{8}$, CO$_{2}$); short lived species also exhibit steeper gradients in their winter polar altitude profiles. Between January 2010 and September 2011, the mixing ratios of some molecules (HC$_{3}$N, C$_{6}$H$_{6}$, C$_{3}$H$_{4}$, C$_{4}$H$_{2}$, C$_{2}$H$_{4}$, HCN) increased by factors of 500 to 1000 between 0.02 mbar ($\sim$365 km) and 0.001 mbar ($\sim$500 km) as the south pole moved into winter \citep{Vinatier:2015}. By 2012, the hemispheric asymmetry in the abundances of HC$_{3}$N and C$_{6}$H$_{6}$ had flipped from the trend observed in 2010 at the poles \citep{Coustenis:2016}.

GCMs are generally consistent with the observed latitudinal and seasonal trends in mixing ratios, but they usually overestimate the latitudinal extent of enhancements at the winter pole and do not reproduce an observed depletion in species like C$_{2}$H$_{2}$, HCN, and C$_{4}$H$_{2}$ at 0.01 mbar and 50$^{\circ}$N, which may result from residual circulation \citep{Teanby:2008}. 

Comparisons between Cassini CIRS and Voyager IRIS measurements indicate that for most molecules at similar latitudes and seasons the observed abundances are consistent with no inter-annual variability \citep{Coustenis:2013, Coustenis:2016}; small enhancements relative to Voyager are observed for HC$_{3}$N at 70$^{\circ}$N \citep{Coustenis:2016} and C$_{4}$H$_{2}$ and C$_{3}$H$_{4}$ at 50$^{\circ}$N \citep{Coustenis:2013}, which may be explained by the observations not perfectly overlapping in season at a given latitude. 

In addition to mixing ratio variations, seasonal variations in the temperature profile are also observed. The winter polar stratospheric temperature structure is consistent with subsidence and reminiscent of Earth's winter pole. In the lower stratosphere, temperatures are at least 25 K colder than the equator \citep{Achterberg:2008}. However, higher in the atmosphere the stratopause is $\sim$20 K warmer than at the equator and is about two scale heights higher; the temperatures exceed 200 K and it is the warmest region of Titan's atmosphere \citep{Flasar:2005, Achterberg:2008, Achterberg:2011, Vinatier:2015}. The heating likely results from adiabatic heating from subsidence over the winter pole and has been predicted by GCMs (see e.g., \citet{Rannou:2005, Crespin:2008, Newman:2011}). Additionally, as \citet{Achterberg:2008} point out, the combination of higher polar altitudes in permanent sunlight and the increased aerosol abundance at the winter pole may result in additional heating from stratospheric aerosols. This temperature structure was first observed during northern hemisphere winter, but similar behavior has now been observed at the south pole, with temperatures of $\sim$185 K seen at the stratopause \citep{Achterberg:2011, Vinatier:2015}. Deeper in the atmosphere, temperature decreases attributed to radiative cooling have been observed as the south pole moves into winter; from 2010 to 2014, the temperature at 0.1 mbar at 70$^{\circ}$S decreased by 40 K \citep{Coustenis:2016}, with very dramatic temperature changes beginning in February 2012 \citep{Vinatier:2015}.

\citet{Teanby:2010b} found that there were no significant changes from 40$^{\circ}$S to 30$^{\circ}$N in composition from 2004 to 2010 and \citet{Coustenis:2016} found that from 2010 to 2014 the mixing ratios did not change significantly at 50$^{\circ}$N and until 2013 remained higher than the mixing ratios at 50$^{\circ}$S. At the mid latitudes, \citet{Coustenis:2016} found that the temperature does not vary significantly in the lower stratosphere (below 0.5 mbar) but higher up in the stratosphere it warmed by 10 K between 2010 and 2014 at 50$^{\circ}$N while at 50$^{\circ}$S it initially decreased from 2010 to 2012 before warming 8 to 10 K.

As a pole moves into summer, the downwelling decreases, resulting in a decrease in adiabatic heating and a decrease in temperature higher in the atmosphere, while the deeper atmosphere warms as the photon flux increases. The gas mixing ratio enhancements also begin to dissipate. At the north pole by 2014, the mixing ratios of some species were beginning to decrease and a 6 K increase in temperature was observed from February to September of 2014 \citep{Coustenis:2016}.

Models predict that for a brief period a few years after equinox there are two equator to pole Hadley cells as the main pole to pole circulation switches directions (see e.g., \citet{Hourdin:1995, Hourdin:2004, Newman:2011, Lebonnois:2012}). \citet{Vinatier:2015} analyzed limb profiles (covering 150 to 500 km) from 2006 to 2013 and reported the first observational evidence of 2 short lived equator to pole Hadley cells in analyses of gases, aerosols, and temperature from Cassini CIRS. Their results, including increased aerosol abundance at the south pole and temperature changes, are compatible with two cells downwelling at the poles from at least January 2010 to at least June 2010. By June 2011, it appears that the circulation had reversed and was again a single Hadley cell with the downwelling branch now located at the south pole \citep{Vinatier:2015}. Models also predict that the upwelling branches near the equator should cause a small temperature decrease from adiabatic cooling, but the temperature change may be too small to observe \citep{Vinatier:2015}. CIRS observations in January 2012 indicate that there was an anomalously low aerosol concentration at the equator, which, if real, may indicate upwelling of air that is depleted in particles \citep{Vinatier:2015}.

\subsection{Detached haze layer}

\begin{figure}
{\includegraphics[width=0.9\textwidth]{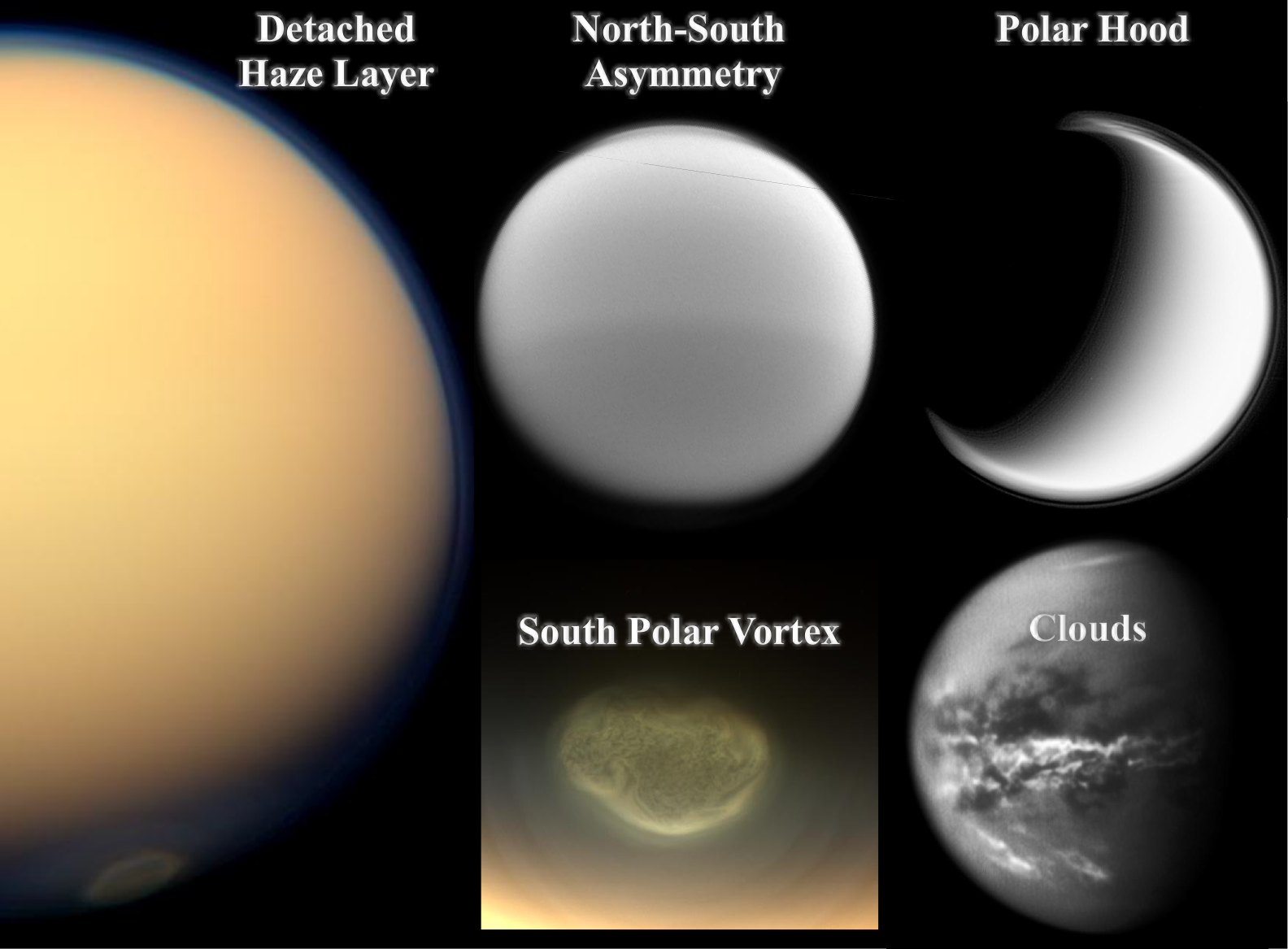}}
\caption{Shown here are examples of some of Titan's distinctive atmospheric features. At the left, a natural color view of Titan from 2012 (PIA14925; July 25, 2012) showing the detached haze layer and the newly formed vortex at the south pole (shown also bottom middle, PIA14919; June 27, 2012). The north-south asymmetry in the haze is shown top middle (PIA14610; January 31, 2012). Titan's north polar hood stands out in the image shown top right (PIA08137; January 27, 2006). At the bottom right, extensive methane clouds at a number of different latitudes, including the equator, show the post-equinox shift in cloud location (PIA12810; October 18, 2010). All image credits are NASA/JPL/Space Science Institute.}
\label{fig:atmosphere}
\end{figure}

The origin of the detached haze layer, shown in Figure \ref{fig:atmosphere}, is not yet well understood. As mentioned above, before the discovery of very large ions in Titan's thermosphere, we assumed that the haze particles were produced at roughly the same altitude as the observed haze layers and that the detached haze layer was either the result of photochemistry \citep{Chassefiere:1995, Clarke:1997, Rannou:2000} or dynamics in which the detached haze layer is the result of large scale circulation in which particles are transported by meridional circulation to the winter pole where they grow, settle, and mix in the stratosphere. In the summer hemisphere, particles are transported upward to maintain the detached haze layer and then transported horizontally by winds (see e.g.\ \citet{Toon:1992, Rannou:2002, Rannou:2003, Rannou:2004}). The GCM of \citet{Lebonnois:2012} accordingly predicts that the detached haze layer changes altitude because of the transient migration of the ascending branch of the Hadley cell. The Imaging Science Subsystem (ISS) and UVIS have detected a detached haze layer at 500 km, along with less pronounced layers above the main detached haze layer \citep{Porco:2005, Liang:2007, Koskinen:2011}. This location is about 150 km higher than the detached haze layer observed by Voyager as shown in Figure \ref{fig:haze}. A temperature minimum in the region of the detached haze layer \citep{Fulchignoni:2005, Sicardy:2006} has led to the suggestion that the detached layer is a result of condensation in the atmosphere \citep{Liang:2007}. However, subsequent analyses by \citet{Lavvas:2009} show that the detached haze layer is coincident with, and likely the cause of, a temperature maximum thus ruling out condensation as a source mechanism. \citet{Lavvas:2009} suggested that the detached haze layer may result from the transition from 40 nm monomers to larger aggregates in the atmosphere, which would cause a decrease in the number density of aerosols. They also demonstrated that the mass flux out of the detached haze layer is sufficient to generate the main haze layer. Their explanation lacks a seasonal component, but the dynamical explanations do not reproduce some of the observations such as the spatial continuity of the detached haze layer at the summer pole \citep{West:2014}. Given these difficulties, it seems likely that both dynamics and microphysics play a role in determining the structure and temporal evolution of the detached haze layer.

The layered structure observed above the detached haze layer may result from gravity waves \citep{Strobel:2006, Koskinen:2011}. Gravity waves have also been suggested as an explanation for similar layers observed in Pluto's atmospheric haze \citep{Gladstone:2016}. 

From 2007 to 2010, the location of the detached haze layer dropped from 500 km to 380 km \citep{West:2011} returning to the same altitude at which it was observed during the Voyager era at the same point in Titan's year. The apparent seasonal evolution in the presence and location of the detached haze layer strengthens the argument that atmospheric dynamics play a role in its origin. The drop in altitude was most rapid at equinox. Starting in late 2012 the detached haze layer was not detectable until early 2016 when it reappeared, with very low contrast, near 500 km \citep{West:2016b}.

\citet{Cours:2011} find that the detached haze layer is composed of two populations of particles: particles less than 50 nm (81-87\% small aerosols by mass), which are primarily responsible for absorption due to particles in that region of the atmosphere, and larger particles, although they are less abundant, which are responsible for 90\% of the scattering at UV and visible wavelengths. The larger particles are too large to have formed at higher altitudes and therefore must have been transported from a reservoir of larger particles deeper in the atmosphere \citep{Cours:2011}. \citet{Larson:2015} find that the location of the detached haze layer represents the location where the upward velocity from dynamics equals the fall velocity due to gravity and thus represents the maximum combination of particles from above and below. They find that it varies seasonally because of seasonal changes in the vertical wind speeds and that it is relatively uniform with latitude because meridional motions are much faster than vertical motions. However, they predicted it would reappear earlier than it has. Note that a few Titan GCMs qualitatively reproduce the seasonal variation in presence and altitude variation of the detached haze layer \citep{Rannou:2002, Rannou:2004, Lebonnois:2012, Larson:2015} however the observed timing and locations have not been reproduced by models. The inability to reproduce the altitude is likely due to limitations imposed by the top of the model atmospheres, which is generally near the highest observed altitude of the detached haze layer \citep{Lebonnois:2012, Larson:2015}.  

\subsection{North-south haze asymmetry}
Titan's north-south haze asymmetry, shown in Figure \ref{fig:atmosphere}, has been observed from ground- \citep{Lockwood:1977, Lockwood:1979, Gibbard:2004b, Lockwood:2009} and space-based telescopes \citep{Lorenz:1997, Lorenz:2004, Karkoschka:2016b}, Pioneer 11 \citep{Tomasko:1982}, Voyager 1 \citep{Sromovsky:1981}, and Cassini (see e.g., \citet{Rannou:2010, Barnes:2013b, Hirtzig:2013}). The winter hemisphere is darker at short wavelengths and brighter in the near IR due to an increase in haze opacity (see e.g., \citet{Toon:1992, Cours:2010, Larson:2014}), which is consistent with other measurements that show that haze is more abundant in the winter hemisphere (see e.g., \citet{Rannou:2010}).

HST observations from 1992 to 1995 showed that the Voyager asymmetry had flipped hemispheres \citep{Lorenz:1997} indicating that it is a seasonal asymmetry. The ground-based observations of \citet{Lockwood:2009} suggest the reversal happened sometime in the mid 1980s. \citet{Penteado:2010} demonstrated that variations in the aerosol optical depth above 80 km could be responsible for the change, but are unable to ascertain if it is a change in composition, size, and/or vertical distribution. Analysis of HST STIS data from 1997 to 2004 using principal component analysis \citep{Karkoschka:2016b} shows that there are two principal components resulting in the asymmetry: one below 80 km and one above 150 km, which they attribute to condensation and dynamics, respectively. Between 80 and 150 km there was no change. The high altitudes were symmetric about equator in 2001 and then switched; at low altitudes the change happened later than at high altitudes and was latitude dependent \citep{Karkoschka:2016b}, which is consistent with \citet{Lorenz:2004}. \citet{Hirtzig:2013} observed a decline in the asymmetry but it had not flipped by mid-2010; \citet{Karkoschka:2016b} predicted the next reversal might occur throughout 2017 and 2018. Inspection of publicly available Cassini ISS images in the MT3 (889 nm) filter suggests that the reversal is currently happening (late 2016). The long time baseline (greater than a Titan year) of the observations of \citet{Lockwood:2009} allowed them to demonstrate inter-annual variability in Titan's brightness, indicating that Titan not only exhibits seasonal changes, but also annual variability.  

\subsection{Polar hood}

In addition to the detached haze layer and main haze layer, which although temporally and spatially varying, are global in extent, Titan's winter pole also has a layered haze structure between the main haze layer and the detached haze layer that resembles a polar hood, as shown in Figure \ref{fig:atmosphere}. The polar hood is likely the result of transport of haze particles and downwelling at the winter pole \citep{Rannou:2002, Rannou:2004, Lorenz:2006c}, which is consistent with its presence at the north pole during the Voyager Era and disappearance from the south pole in 2002-2003 \citep{Lorenz:2006c}.

\section{Clouds, rain, and Titan's methane cycle \label{Sect:clouds}}

Much like Earth, Titan exhibits a number of types of clouds resulting from different formation processes and conditions. Unlike Earth, Titan's clouds form from a number of different volatiles. \citet{Tsai:2012} divided clouds into 3 categories: convective methane clouds \citep{Griffith:1998, Griffith:2000, Brown:2002, Roe:2002, Gibbard:2004, Griffith:2005, Roe:2005, Schaller:2006, Schaller:2006b, Griffith:2009, Rodriguez:2009, Schaller:2009, Turtle:2009, Brown:2010, Rodriguez:2011, Turtle:2011}, stratiform ethane clouds \citep{Griffith:2006, Brown:2010, LeMouelic:2012}, and high altitude cirrus clouds (HCN, HC$_{3}$N, etc.) \citep{Samuelson:2007, Anderson:2010, Anderson:2011}. In addition to those clouds, there is currently a large, high altitude HCN ice cloud, first observed in May 2012, that formed in the polar vortex at the south pole \citep{dekok:2014}. Figure \ref{fig:atmosphere} shows the polar HCN cloud and an example of methane clouds. Although there have been reports of detection of fog, \citet{Charnay:2012} argue that those observations are more likely boundary layer clouds. In general, convective methane clouds are observed at the summer pole and mid-latitudes, while the other types of clouds are observed at the winter pole.

\subsection{Convective methane storms}
The GCMS aboard the Huygens Probe measured a methane relative humidity of approximately 50\% at the surface \citep{Niemann:2005, Niemann:2010}. However, cloud and thunderstorm models indicate that at least 60\% surface relative humidity is required for convective cloud formation \citep{Lorenz:2005} and 80\% is required for thunderstorms capable of producing substantial rainfall \citep{Hueso:2006}. It has therefore been argued that there is not currently enough methane at the equator to produce rainfall necessary to form the channels observed at the Huygens landing site \citep{Griffith:2008}. However, low latitude storms have been observed \citep{Schaller:2009}, including one that resulted in extensive alteration of the surface, presumably caused by large amounts of methane rainfall \citep{Turtle:2011b}. These convective storms may produce fast surface winds that increase the lifetime of the storm and enhance precipitation \citep{Hueso:2006, Rafkin:2015, Charnay:2015}.

The methane and temperature profiles at the Huygens landing site indicate that Titan's tropical atmosphere is highly stable; convection is weak and convective systems lie near or below 26 km \citep{Barth:2007, Griffith:2008}. Close to the surface (below 2 km) the temperature profile follows the dry lapse rate \citep{Griffith:2008} and parcels of air from the surface need to reach the lifting condensation level (LCL) of 5 km before condensation will occur \citep{Griffith:2008}. The level of free convection (LFC) is around 9 km and the level of neutral buoyancy (LNB) is around 24 km \citep{Griffith:2008}. Many of the observed clouds that form and dissipate on short time scales are at altitudes near the LNB \citep{Griffith:2000, Griffith:2005, Griffith:2009, Adamkovics:2010} indicating that they are convective \citep{Griffith:2000, Griffith:2005, Roe:2012b}. Interestingly, the clouds at higher latitudes also reach higher altitudes (up to 40 km), potentially indicating that the methane humidity profile varies latitudinally \citep{Griffith:2005, Griffith:2009, Adamkovics:2010, Griffith:2014} as discussed in Section \ref{sec:methane}. 

Dynamical models of Titan's atmosphere indicate that there are four ways to produce clouds: updrafts (including the ascending branch of a Hadley cell) \citep{Rannou:2006, Mitchell:2006, Mitchell:2008, Mitchell:2009b, Schneider:2012}, solar forcing at the summer poles, wave activity, and downwelling of photochemical species \citep{Griffith:2014}. The clustering of clouds at 40$^{\circ}$S during southern summer has been explained by either the location of slantwise convection \citep{Rannou:2006} or the ascending branch of the Hadley cell \citep{Mitchell:2006}. Planetary waves \citep{Schaller:2009}, Kelvin waves \citep{Mitchell:2011, Turtle:2011b}, and atmospheric tides \citep{Rodriguez:2009, Rodriguez:2011} have all been used to explain the location, timing, and/or morphology of observed clouds. Clouds produced by waves are particularly important for the tropics near equinox \citep{Mitchell:2011, Schneider:2012}. \citet{Barth:2004} predicted that optically thick clouds are most likely to form at the poles. Mountains, resulting in orographic clouds \citep{Barth:2010}, and lakes and seas, resulting in sea breeze clouds \citep{Brown:2009}, may also play an important role in cloud formation on Titan.

It is generally assumed that haze particles produced higher in the atmosphere act as cloud condensation nuclei (see e.g., \citet{Griffith:2006, Bauerecker:2009}); liquid and solid CH$_{4}$ nucleate easily on Titan haze analogs \citep{Curtis:2008} but CH$_{4}$-N$_{2}$ binary nucleation may also play an important role in cloud formation \citep{Wang:2010, Tsai:2012}.

Since clouds were first discovered using ground-based observations \citep{Griffith:1998, Brown:2002, Roe:2002}, observations of clouds have played an important role in improving our understanding of the dynamics of Titan's atmosphere. In addition to providing measurements of wind speeds, the powerful combination of long-term ground-based cloud monitoring campaigns (e.g., \citet{Roe:2005, Schaller:2006, Schaller:2006b, Schaller:2009, Adamkovics:2010}) and observations by Cassini (e.g., \citet{Porco:2005, Turtle:2009, Brown:2010, Rodriguez:2011, Turtle:2011}) allow us to track seasonal changes in cloud cover, including location, extent, and duration, which provide important constraints on the global circulation of Titan's atmosphere and a test for GCMs. Clouds can also help us understand atmospheric waves and are particularly important for providing information about the troposphere, which is difficult to obtain from other types of remote sensing because of the large optical depth. The story of clouds on Titan is a particularly powerful demonstration of the synergy of ground and spacecraft based observations; Ground based observers have more consistent measurements, over a longer time baseline but are unable to detect smaller clouds. Cassini measurements can detect much smaller cloud features, but the frequency, consistency, and length of time baseline for the observations is better from the Earth;  \citet{Schaller:2009} found a storm from ground-based observations that was largely missed by Cassini.

Large cloud systems (covering up to $\sim$10\% of the disk) seem to occur every 3 to 18 months \citep{Roe:2012b}. These large storms are capable of triggering Rossby waves, which result in formation of clouds elsewhere in the atmosphere \citep{Schaller:2009}. Despite the presence of large storms, lightning has not been detected in measurements of Titan's atmosphere from Cassini-Huygens \citep{Beghin:2007, Fischer:2007, Fischer:2011}.

Cassini's RADAR should be capable of detecting raindrops, but a dedicated cloud backscatter observation during T30 did not detect droplets \citep{Lorenz:2008c}. Although raindrops have not been directly detected, a handful of probable rainfall events have been identified based on observed surface changes, some of which have been tied to specific observed storms \citep{Turtle:2009, Turtle:2011b, Dalba:2012, Barnes:2013}. On Titan, precipitation may be supply limited; if that is the case, we would expect frequent large storms during northern summer \citep{Mitchell:2012}. Since the average mixing ratio and atmosperhic column mass of methane in Titan's atmosphere is much higher than that of water on Earth, when storms do produce rain, they are capable of significant amounts of precipitation \citep{Hueso:2006, Barth:2006, Barth:2007, Barth:2010}. Estimates of the average precipitation rate of methane to Titan's surface generally fall between 0.001 and 0.5 cm/terrestrial-year \citep{Barth:2006, Tokano:2006}.

The model of \citet{Mitchell:2011} found that the ``arrow storm" \citep{Turtle:2011b} was caused by an equatorially trapped Kelvin wave, which is one of the dominant modes of dynamics in Titan's atmosphere. These waves result in large-scale storms (over 1000 km scale that produces several centimeters of precipitation, which is 20 times the average rate) indicating that it may rain more frequently in the equatorial region than previously thought.

\subsection{Seasonal evolution of convective clouds}
These storms tend to appear in the regions of maximum solar insolation, which results in seasonal variation in storm locations (see e.g., \citet{Brown:2010, Turtle:2011}) as the region of upwelling, similar to the intertropical convergence zone (ITCZ) on Earth, moves from pole to equator to pole. Near southern summer solstice (2002) most of the observed storms were near the south pole. As Titan moved through equinox (2009), clouds appeared at both poles. From 2013 to 2015, very few clouds were detected from Cassini or ground-based monitoring campaigns \citep{Rodriguez:2014, Corlies:2015, Turtle:2016b}, indicating that cloud activity may be less frequent post-equinox. In 2016, small clouds started appearing more consistently; however, the 2016 clouds are still much smaller than some of the clouds observed near equinox at high northern latitudes \citep{Cassini:2016, Turtle:2016b}. The timing of clouds appearing at the north pole was not predicted by models and is an area of ongoing work.

In general GCMs reproduce the observed trends (see e.g., \citet{Mitchell:2006, Mitchell:2009b, Schneider:2012, Lora:2015b, Newman:2016}), although the presence and/or timing of mid-latitude clouds is not seen in every model leading \citet{Lora:2015b} to argue that the clouds observed at mid-latitudes are either not convective in origin or are generated by another mechanism like orographic gravity waves.  

\subsection{Ice clouds}

Since the trace constituents in the atmosphere are produced by photochemistry, their abundances tend to decrease with decreasing altitude. Cloud formation from these species therefore requires transport of air parcels downward, from regions where the mixing ratio is higher. A number of opacity sources in Titan's atmosphere are consistent with particles that form from subsidence; some have been connected with specific species such as HC$_{3}$N \citep{Coustenis:1999, Khanna:2005, Anderson:2010}, HCN \citep{Samuelson:1997, Clark:2010, dekok:2014}, C$_{4}$N$_{2}$ \citep{Khanna:1987, Samuelson:1997, Coustenis:1999}, and C$_{2}$H$_{6}$ \citep{Griffith:2006, Anderson:2011}, while others remain unidentified. In general, particles formed are on the order of a few microns in size (see e.g., \citet{Coustenis:1999, Mayo:2005, Samuelson:2007, Griffith:2009, Anderson:2011, dekok:2014}).

In general nitriles condense at a higher altitude than hydrocarbons (see e.g., \citet{Mayo:2005, Anderson:2011, Lavvas:2011}. A feature at 160 cm$^{-1}$ observed at 90 km has been attributed to a mix of HCN and HC$_{3}$N \citep{Anderson:2011}, a region of the atmosphere where HCN and HC$_{3}$N are expected to condense (see e.g., \citet{Anderson:2011, Lavvas:2011}). This feature is observed from 58$^{\circ}$S to 85$^{\circ}$N (covering the full latitude range analyzed), is found at higher altitudes at higher latitudes, and the abundance increases from south to north \citep{Samuelson:2007, Anderson:2011}.

No optically thick clouds were detected in DISR measurements \citep{Tomasko:2005, Doose:2016}, however thin layers observed at 7, 11, and 21 km have been attributed to some combination of phase transitions, gas adsorption, and atmospheric mixing, resulting in a particle size increase of 5-10\% \citep{Tomasko:2005, Tokano:2006, Karkoschka:2009}. The layer at 21 km was also likely detected by HASI/PWA, which detected extremely low frequency noise possibly resulting from aerosols impacting the electrodes of the antenna \citep{Beghin:2007}. 

\citet{Lavvas:2011c} examined condensation processes at the Huygens landing site. Above 80 km, the DISR observations can be fit using only a model of particle coagulation and sedimentation. However, below 80 km condensation must be invoked to fit the observations. Between 30 and 80 km the aerosol opacity is dominated by a coating of HCN. Although HCN is removed from the atmosphere by this process, deposition of galactic cosmic rays (GCRs) results in production of HCN and replenishes the HCN that is lost allowing for continuous condensation onto aerosols. Below 30 km condensation of methane onto the particles explains their optical properties. They also note that while ethane clouds are likely to form in the troposphere, the droplets would evaporate before they reach the surface. This may also happen with methane clouds \citep{Lavvas:2011c}. 
 
\subsubsection{The ``Haystack'' (220 cm$^{-1}$ feature)}

A far infrared feature, centered at 220 cm$^{-1}$ and sometimes referred to as ``the haystack'' \citep{West:2014}, was first observed in the data obtained by Voyager IRIS \citep{Kunde:1981, Coustenis:1999}. It was present at northern high latitudes from 2004 to 2012 and observed frequently by CIRS; it is most likely due to the presence of solid particles from 80 to 150 km altitude (see e.g., \citet{Flasar:2004, dekok:2007b, dekok:2008, Samuelson:2007, Jennings:2012b, Jennings:2015}). From 2004 to 2012, the intensity of the feature decreased by a factor of four indicating seasonally varying processes (strength of downwelling, solar heating, photochemistry) are responsible for its formation \citep{Jennings:2012b}. The variation in strength of the feature is similar to that of HC$_{3}$N leading \citet{Jennings:2012b} to argue that it has a nitrile origin. \citet{Jennings:2012} reported the first observation of the feature in spectra of the south pole taken in July 2012 (it was not present in February 2012), indicating that it is a product of processes that occur at the winter pole. From the time it appeared the intensity has increased rapidly, doubling every year \citep{Jennings:2015} as the temperature has decreased at the south pole \citep{Teanby:2012, dekok:2014, Vinatier:2015, Coustenis:2016}. The feature is offset 4 degrees from the pole, consistent with the other offsets as described in Section \ref{sec:dynamics}, and as of December 2013 it was confined to a ring south of 80$^{\circ}$S. Nitriles and complex hydrocarbon gases are also offset but with a central peak at the pole, a minimum at 80$^{\circ}$S and secondary maximum at 70$^{\circ}$S \citep{Jennings:2015}. By September 2014, the ring structure in the gases dissipated \citep{Coustenis:2016}.

\subsubsection{Dicyanoacetylene (C$_{4}$N$_{2}$)}

Although C$_{4}$N$_{2}$ has not been detected in Titan's atmosphere \citep{Samuelson:1997, Jolly:2015}, a feature at 478 cm$^{-1}$ has been attributed to a dicyanoacetylene ice cloud at high latitudes during northern winter below 90 km \citep{Khanna:1987, Samuelson:1997, Coustenis:1999}. \citet{Anderson:2016} argue that the C$_{4}$N$_{2}$ is produced through solid-state photochemistry of HCN-HC$_{3}$N mixtures, which would explain how C$_{4}$N$_{2}$ could exist in the condensed phase and be undetectable in the gas phase. 

\subsubsection{Ethane (C$_{2}$H$_{6}$)}

In addition to the large methane storms observed in Titan's atmosphere, an extensive polar ethane cloud has also been detected at an altitude 30 to 65 km poleward of 50$^{\circ}$N \citep{Griffith:2006, Rannou:2010}, an altitude region where \citet{Barth:2003} demonstrated ethane ice clouds are stable. The cloud was present when Cassini arrived at Titan (from 51$^{\circ}$N to 68$^{\circ}$N, the highest latitude illuminated at the time) and persisted until 2009 when it began to dissipate as Titan approached spring equinox \citep{Brown:2010, LeMouelic:2012}. It appears to be the result of the descending branch of winter circulation and the extent and altitude is broadly consistent with GCMs \citep{Rannou:2006, Rannou:2012}. Since the cloud appears to be a direct result of dynamical processes, the location, extent, and variation with season provides a useful test for models of Titan's atmospheric circulation. Interestingly, after the breakup of the cloud near equinox, the region north of 62$^{\circ}$N is depleted in particles \citep{Rannou:2012}. \citet{Anderson:2014} argue that CH$_{4}$ ice particles formed from subsidence may contribute significantly to the opacity of this extensive cloud because the winter pole is cold enough to condense CH$_{4}$ ice. 

\subsubsection{Hydrogen cyanide (HCN)}

In 2012, at the location of what was previously a maximum in the atmospheric temperature profile at the south pole, a large cloud was observed in Cassini ISS data at an altitude of 300 km \citep{dekok:2014, West:2016}. \citet{dekok:2014} used Cassini VIMS data to identify the cloud as a HCN ice cloud with particles of approximately 1 $\mu$m. Cassini CIRS measurements indicate much more rapid cooling than GCMs predicted and the temperature would have to be $\sim$100 K lower than predicted for HCN ice formation to occur \citep{dekok:2014}. The cloud has distinct boundaries, mottled texture, and is more yellow and brighter than the surrounding haze \citep{West:2016}. It is rotating offset 4.5$^{\circ}$ from the pole, which is consistent with \citet{Achterberg:2008b}, and exhibits zonal wind speeds of 87 to 89 m/s \citep{West:2016}. \citet{West:2016} point out that the morphology is reminiscent of open cell convection on Earth. They suggest that on Titan it may form as the pole approaches winter and begins cooling to space, the radiative relaxation time decreases with increasing altitude and the condensible species (HCN) comes from above creating a destabilization that results in the texture. Interestingly, the cloud formed at the same altitude as the detached haze layer but the relationship, if any, between these two features remains to be explored \citep{West:2016}. 

\subsection{Methane cycle}
Although there are numerous analogies between Earth's hydrological and Titan's volatile cycle, it is important to remember that unlike water on Earth, methane can be transported through the cold trap at the tropopause, which allows it to play an important role in the radiation balance and chemistry in other regions of the atmosphere including allowing it to be destroyed in the upper atmosphere \citep{Roe:2012b}. Thus although methane does cycle back and forth between the atmosphere and the surface, it is also lost in significant amounts from the atmosphere and the cycle is only stable over geologic timescales if there is a source of methane. Another difference is that on Titan, most of the methane is in the atmosphere rather than on the surface. If all of the methane currently in Titan's atmosphere condensed onto the surface, it would form a global layer 4-5 m deep \citep{Penteado:2010}. For a more detailed discussion of clouds and Titan's methane cycle, see the excellent reviews by \citet{Lunine:2008} and \citet{Roe:2012b}.

Over the years a number of possible methane sources have been suggested including outgassing of a deep reservoir \citep{Tobie:2005, Tobie:2006}, interior chemistry involving organic material \citep{Atreya:2006}, seepage of methane from a shallow subsurface reservoir \citep{Soderblom:2007, Hayes:2008, Griffith:2012}, and cryovolcanism. Our lack of understanding about the near surface and subsurface reservoirs of methane and other hydrocarbons limits our ability to fully understand the methane cycle on Titan. Aside from the information from the Huygen's landing site about volatiles in the surface, discussed below, there is other evidence that replenishment from the subsurface may play an important role in Titan's volatile budget. \citet{Neish:2014} find that Titan's crater distribution indicates widespread wetlands of liquid hydrocarbons at low elevations over much of geologic time. As the north pole moves into summer, the warming is lagging predictions potentially indicating that the north polar region is saturated with liquid \citep{Jennings:2016, Legall:2016}. Additionally, there is some evidence of low latitude lakes, which are not stable and would require subsurface replenishment \citep{Griffith:2012}.

One additional outstanding issue with Titan's methane is whether or not it is escaping from Titan's upper atmosphere. The measured profile in the thermosphere defies explanation. If methane is escaping, it represents only a small fraction of the total methane budget and therefore does not significantly impact the methane lifetime mentioned above \citep{Strobel:2014}.

\section{Connection with surface \label{Sect:surface}}

\begin{figure}
{\includegraphics[width=0.9\textwidth]{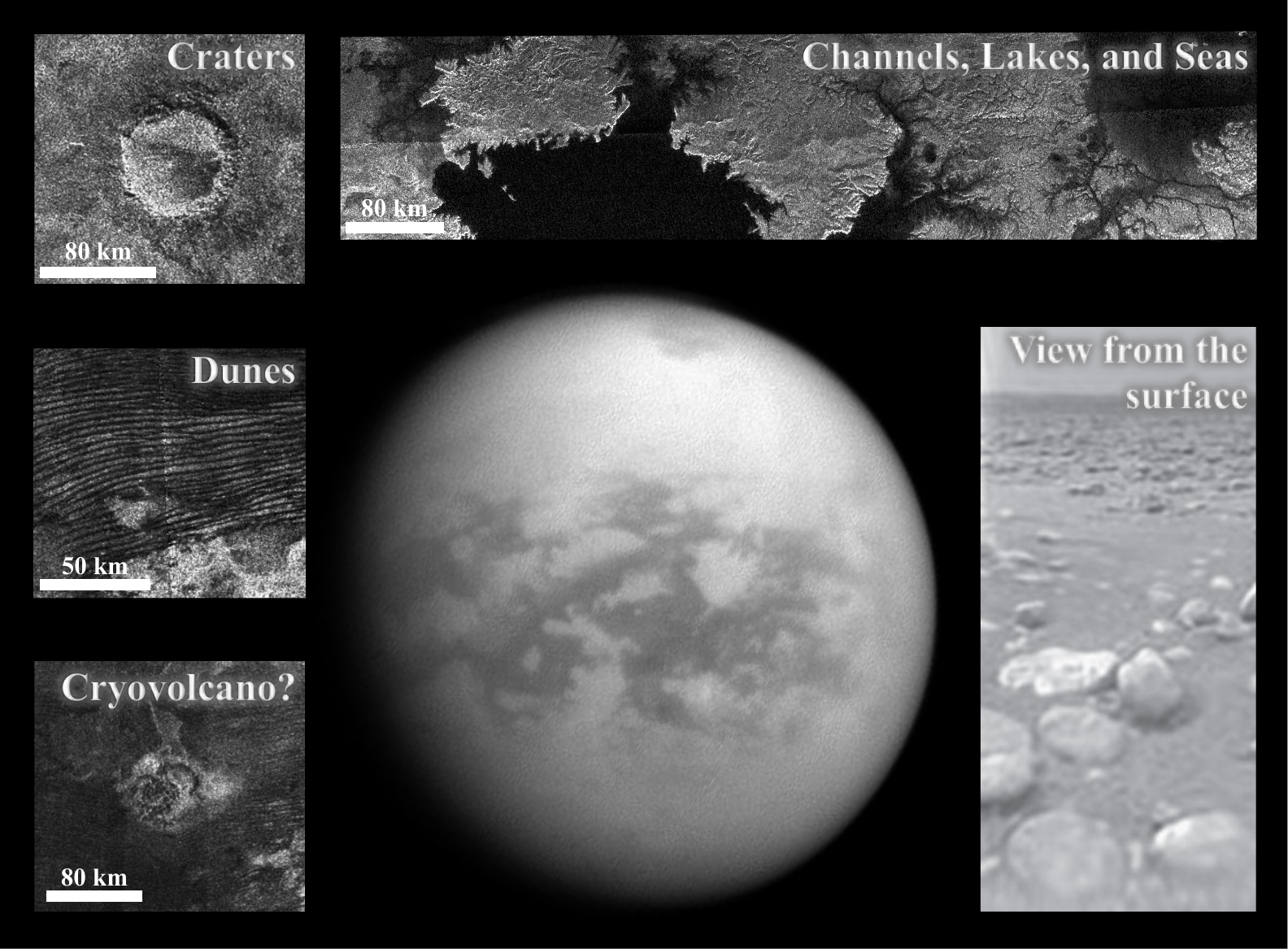}}
\caption{Shown here are examples of some of the surface features discussed in the text. The crater shown (top left) is Sinlap, which has a diameter of 79 km (PIA16638, NASA/JPL-Caltech/ASI/GSFC). An example of Titan's equatorial dunes from the Shangri-la dunes is shown middle left (PIA12037, NASA/JPL-Caltech/ASI). Doom Mons and Sotra Patera, a putative cryovolcanic region, are shown bottom left (PIA09182, NASA/JPL-Caltech/ASI). At the top is Kraken Mare (left) and Ligeia Mare (right) and associated channel systems (PIA09217, NASA/JPL-Caltech/ASI). Shown bottom right is the view from the surface taken by the Descent Imager Spectral Radiometer carried by Huygens. The cobbles in the foreground are 10 to 15 cm (PIA06440, ESA/NASA/JPL/University of Arizona). The middle image was taken at 938 nm (CB3 filter), which sees down to the surface. The dark equatorial regions are Titan's expansive dune fields, while the dark regions at the north pole are lakes and seas (PIA14584, NASA/JPL-Caltech/Space Science Institute).}
\label{fig:surface}
\end{figure}

Titan's surface possesses a variety of environments including lakes, mountains, and dunes as shown in Figure \ref{fig:surface}. As discussed below, the surface has been shaped by aeolian and fluvial processes and the connection between Titan's surface and atmosphere allows us to test theories developed for Earth in completely different environment. 

\subsection{Surface composition}

Despite analysis of data obtained \emph{in situ} by the GCMS carried by Huygens and remote sensing data from DISR, VIMS, and RADAR, the composition of Titan's surface remains poorly understood. Chemical and physical processes occurring in the atmosphere result in the formation of organic liquids and solids that are eventually deposited on the surface; the surface composition is uniquely tied to its atmosphere. Based on Titan's bulk density ($\sim$1.88 g/cm$^{3}$, \citet{Jacobson:2006}) water ice makes up a large fraction of Titan's mass and is therefore another likely candidate surface material. Evidence for water ice on Titan's surface was first obtained by ground-based telescopes \citep{Griffith:1991, Griffith:2003} and has also been seen in VIMS \citep{McCord:2006, Griffith:2012, Hayne:2014} and DISR data \citep{Tomasko:2005, Rannou:2016}. Others have argued that these features can be explained by other hydrocarbons or nitriles \citep{Clark:2010}. Other suggested surface constituents from interpretation of VIMS data include CO$_{2}$ \citep{Barnes:2005, McCord:2007}, C$_{6}$H$_{6}$, C$_{2}$H$_{6}$, and HC$_{3}$N \citep{Clark:2010}. Our ability to constrain the composition of Titan's surface is fundamentally limited by the resolution of VIMS and the small wavelength windows where it is possible to observe the surface from orbit. VIMS data can be used to identify distinct units on the surface \citep{Barnes:2007b, Soderblom:2007}, but conclusive identification of the composition of the units is probably not possible without another mission given the constraints of the instrument. For example, Xanadu, one of Titan's most prominent features and the first surface feature seen from earth \citep{Lemmon:1993, Smith:1996}, is a ``bright'' terrain unit. Other units include ``dark brown'', which seems to be dominated by dunes, and ``dark blue'', which seems to be richer in water ice than the dark brown unit \citep{Barnes:2007b}. \citet{Neish:2015} examined spectra of Titan's craters as a function of degradation state and concluded that impacts expose water ice and organics and then rainfall removes the soluble organics, leaving behind water ice and insoluble organics.

Measurements from Cassini RADAR provide constraints on the dielectric constant of the surface and \citet{Janssen:2016} found that the radar bright regions ($\sim$10\% of Titan's surface) are consistent with near subsurface water ice, while the radar dark regions are consistent with organics. The orbital measurements agree with measurements made by HASI/PWA of the surface at the Huygens landing site, which found a dielectric constant of 2.5 $\pm$ 0.3 and a conductivity of 1.2 $\pm$ 0.6 nS/m; values consistent with photochemically produced organic material in the first meter of the surface \citep{Grard:2006, Hamelin:2016}.

Analyses of measurements taken by the Huygens Probe and assessment of its behavior upon landing indicate that the surface at the Huygens landing site was fine-grained, relatively soft, and possibly covered in a 7 mm thick ``fluffy'' layer \citep{Zarnecki:2005, Lorenz:2006b, Karkoschka:2007, Atkinson:2010, Schroder:2012}. The surface also contained volatiles as indicated by the increase in abundance of CH$_{4}$ after the Huygens Probe landed, most likely resulting from the heating of liquid CH$_{4}$ that was present on the surface\citep{Niemann:2005, Niemann:2010}, attenuation of acoustic signals after landing \citep{Lorenz:2014}, a change in the PWA-MIP observations 11 minutes after landing \citep{Hamelin:2016}, and modeling of DISR spectra \citep{Rannou:2016}. Located in a dry region of Titan, \citet{Williams:2012} interpret the moisture detected at the Huygens landing site as evidence of a recent local precipitation event or the result of recharging from a precipitation event located farther away because the surface should dry relatively rapidly.

Only 11\% of the sunlight incident at the top of Titan's atmosphere reaches the surface \citep{Tomasko:2005}, which means there is very little solar energy available for further chemical processing on the surface. However, the molecules produced in the atmosphere, in particular those that possess triple bonds (like acetylene), carry chemical energy to the surface that could potentially be released \citep{Lunine:2011}. Some laboratory work indicates that GCRs may induce chemistry on the surface \citep{Zhou:2010}, but the majority are deposited near 65 km and more work is required to characterize the surface energetic particle flux.

\subsection{Fluvial transport and lakes and seas}

In addition to methane, a number of products of Titan's atmospheric photochemistry are also liquid at Titan surface conditions; of these ethane is the most abundant. Based on photochemical considerations, \citet{Lunine:1983} argued that the surface might be covered in vast seas or oceans of ethane. Prior to Cassini-Huygens, Earth-based measurements revealed that Titan's surface is heterogeneous \citep{Muhleman:1990, Lemmon:1993, Griffith:1993, Lemmon:1995, Smith:1996}. As the Huygens Probe descended through Titan's atmosphere, images taken by DISR revealed a network of channels flowing into a plain reminiscent of flood plains \citep{Tomasko:2005}. The complex dendritic drainage systems observed are likely indicative of a distributed source and possibly formed by rapid erosion that creates deeply incised valleys \citep{Perron:2006, Soderblom:2007}.  Rounded cobbles at the landing site provided additional evidence of fluvial erosion \citep{Tomasko:2005}. Channel morphology can be used to constrain formation mechanisms. For example, the presence of dry valleys, found only at mid-latitudes, indicates strong episodic events \citep{Langhans:2012}. Well-developed drowned river valleys indicate periods when rising fluid levels outpaced sediment deposition \citep{Stofan:2007}. The spatial distribution of different channel types provides information about spatial variation in the strength and frequency of rainfall. 

Although the Huygens Probe only found evidence of past fluvial activity, RADAR data obtained during T16 revealed numerous lakes and seas near the north pole \citep{Stofan:2007}. For a review of the lakes and seas, see \citet{Hayes:2016}. The surprising ability to use Cassini RADAR to measure the depths of the lakes and seas has resulted in the production of the first extraterrestrial bathymetric maps \citep{Mastrogiuseppe:2014, Legall:2016, Hayes:2016}; Ligeia Mare is up to 160 m deep, while Ontario Lacus is 90 m deep. Titan's lakes and seas cover $\sim$1\% of Titan's surface \citep{Hayes:2008, Hayes:2016} and contain liquids that represent a global ocean depth of about 1 m \citep{Hayes:2016}, significantly less than predicated by photochemical models as discussed in Section \ref{sec:chem}.

Although Titan's entire surface has not been mapped, there appears to be a north-south asymmetry in the distribution of the lakes, which may be the result of Titanian Croll-Milankovitch cycles \citep{Aharonson:2009, Schneider:2012, Lora:2015}. As discussed previously, Titan GCMs agree that methane is efficiently transported away from the equator, drying out the low and mid-latitudes and depositing liquid at the poles \citep{Mitchell:2006, Mitchell:2008, Mitchell:2009b, Schneider:2012, Mitchell:2012, Lora:2014, Lora:2015b}. Although the southern hemisphere appears to be almost entirely devoid of currently filled lakes, there is some geomorphological evidence of paleoseas \citep{Aharonson:2009, Hayes:2011, Moore:2011}, with the most equatorward candidates located at 39$^{\circ}$ and 42$^{\circ}$S \citep{Vixie:2015}.

\citet{Lora:2014} simulated Titan's paleoclimate over the past 42,000 years looking at four characteristic configurations and found that all four result in efficient transport of methane to the poles; the present day configuration and 14 kyr configuration favor transport to the north pole, the 28 kyr configuration favors the south pole, and the 42 kyr configuration is symmetric. This work demonstrates that the lakes may preferentially form at one pole based on orbital configuration, but also shows that the low and mid-latitudes have likely been dry for a long time, and that the asymmetry would reverse with a period of $\sim$125 kyr \citep{Lora:2014}. 

Due to the length of the Cassini mission, it might be possible to detect seasonal changes in lake level by looking for shoreline changes \citep{Mitri:2007}. While some shoreline changes have been reported there are conflicting interpretations  of the data and it remains an open question whether Cassini has detected changes in the liquid level of the lakes and seas \citep{Barnes:2009, Turtle:2009, Moriconi:2010, Hayes:2011, Turtle:2011, Turtle:2011b, Cornet:2012, Sotin:2012, Lucas:2014b, Hayes:2016}. Regardless of whether the lake levels have changed over the past decade, there is evidence for deposits of evaporites on the surface, created by cycles of evaporation and refilling, including around Ontario Lacus \citep{Barnes:2009, Moriconi:2010, Barnes:2011, Cornet:2012b, Mackenzie:2014}. These putative evaporite deposits are anomalously bright at 5 $\mu$m in the VIMS data and $\sim$1\% of the surface exhibits this spectral behavior \citep{Mackenzie:2014}. The signal attributed to evaporites near Ligeia Mare would require it to have covered 9\% greater area indicating lake level changes of unknown timescales \citep{Mackenzie:2014}. The southern paleoseas do not exhibit this signature, indicating burial, erasure, lack of formation in the first place, or some other unknown process \citep{Mackenzie:2014}. The distribution and composition of evaporites contain information about Titan's climate history, but will require composition measurements that cannot be obtained with Cassini.
Measurements \citep{Legall:2016} and recent models \citep{Glein:2013, Tan:2013, Tan:2015} agree that the northern lakes and seas appear to be methane rich, contrary to prior predictions that they should be dominated by ethane \citep{Cordier:2009, Cordier:2012}. Ethane has been detected in Ontario Lacus from VIMS measurements \citep{Brownr:2008} and from RADAR Ontario Lacus appears to contain a greater fraction of ethane than the northern lakes and seas \citep{Hayes:2016}. It is likely that a combination of spatial/temporal variation in precipitation rates (and composition of precipitation), transport \citep{Lorenz:2014c, Tokano:2016}, and thermodynamic considerations will be required to explain the composition of the lakes. Even so, considering that ethane is the most abundant product of Titan's photochemistry, it is surprising that so little liquid ethane has been detected on the surface. \citet{Mousis:2016} showed that ethane could be preferentially pulled into clathrates that are in contact with the lakes and/or an alkanofer, resulting in predominantly methane rich lakes and trapping significant amounts of ethane in clathrates thus explaining the observed lack of liquid ethane on the surface. 

The so-called ``cookie-cutter'' lakes found near the north pole have steep walls and are 100s of meters deep \citep{Hayes:2008, Stiles:2009}, morphology that is consistent with dissolution karst \citep{Cornet:2015}. Since methane does not dissolve water ice but it can dissolve organics like solid acetylene \citep{Glein:2013, Malaska:2014}, the presence of dissolution karst would indicate that there are 100s of meters of organics at the north pole, which has implications for atmospheric chemical and dynamical processes that would result in such a significant deposit of organics.

The lakes and seas represent a region of particular astrobiological interest for Titan. Some fraction of the complex organics produced in the atmosphere will end up in the lakes and seas either directly through sedimentation from the atmosphere (dry removal), washout from rain (wet removal), or from subsequent transport processes once they are deposited on the surface. Some of the compounds are soluble in liquid hydrocarbons, but others are not thus the lakes will partition the organics produced in the atmosphere potentially resulting in subsequent chemistry. \citet{Stevenson:2015} suggested that nitrogen compounds could act as a lipid bilayer in the lakes. In addition to solubility, density also plays a role in the fate of materials in the lakes and seas; ice formed during the winter would likely float \citep{Roe:2012, Hofgartner:2013} and experimental simulations of haze formation indicate the possibility that the haze particles would also float \citep{Horst:2013}. However, hydrocarbon solids are generally more dense than their corresponding liquids and will sink forming deposits on the sea floor \citep{Tokano:2009}. For example, \citet{Legall:2016} see evidence that the bottom of the Ligeia Mare is coated in a layer of nitriles. An improved understanding of the chemistry of the lakes requires experimental investigations and additional measurements from a future mission. 

\subsection{Winds and waves}

The lakes and seas are of particular atmospheric interest because waves can be used to constrain the wind speeds, which as previously mentioned are very difficult to measure directly. For most of the mission, vertical deviations were less than a few mm \citep{Wye:2009, Stephan:2010, Barnes:2011b, Zebker:2014}. The size of the wave depends on wind speed and liquid viscosity, both of which are ultimately determined by the atmosphere. \citet{Lorenz:2012} showed that the 2 m/s maximum wind speeds predicted by GCMs would produce 1 m waves. \citet{Hayes:2013} argue that the lack of detections for most of the mission are not surprising based on GCMs \citep{Schneider:2012, Lora:2015b} as the wind speeds should be below the threshold for generating waves at equinox but should exceed it moving into summer solstice.  As we move into northern summer there have been a few reports of potential wave activity \citep{Barnes:2014, Hofgartner:2014, Hofgartner:2016} and one hopes definitive detections of waves will occur before the end of the mission. 

\subsection{Aeolian transport and dunes} 

Although pre-Cassini-Huygens predictions indicated that Titan is not a favorable location for dune formation \citep{Lorenz:1995}, data from Cassini RADAR revealed widespread regions of equatorial dunes ($\pm$30$^{\circ}$) covering 10-20\% of Titan's surface \citep{Elachi:2005, Lorenz:2006, Radebaugh:2008, Lorenz:2009, Legall:2011, Rodriguez:2014}. The dunes are likely composed of 100-300 $\mu$m particles \citep{Lorenz:2006} and analyses of RADAR and VIMS data indicate that they are composed of pure organic or organic coated materials \citep{McCord:2006, Soderblom:2007b, Barnes:2008, Clark:2010, Legall:2011, Hirtzig:2013, Rodriguez:2014}. The dune particles are much larger than the aerosols measured by DISR near the surface and it is not yet clear whether the dune sands are composed of haze material and if so, how the larger particles formed. \citet{Barnes:2015} discussed a number of possible formation mechanisms including sintering, lithification/erosion, flocculation, and evaporitic precipitation. 

Titan's dunes are eastward propagating longitudinal dunes $\sim$100 m tall, 1 km wide, and on average 30-50 km long \citep{Lorenz:2006, Radebaugh:2008, Barnes:2008, Neish:2010b, Legall:2011}. The observed 1-3 km spacing \citep{Lorenz:2006, Radebaugh:2008, Savage:2014} is approximately the same height as the boundary layer indicating that the dunes are mature, and have therefore stopped growing \citep{Griffith:2008, Lorenz:2010, Lorenz:2014b}. Dune orientation and morphology are determined by wind patterns; given the difficulties in measuring Titan's winds remotely, the dunes provide some of our only constraints on the wind speed and direction at the surface. Although initial GCM wind predictions were inconsistent with the dune orientations \citep{Tokano:2008}, more recent works demonstrate that conditions around equinox, either a reversal of the prevailing winds (to strong westerly (eastward) winds) \citep{Tokano:2010} or the generation of strong, westerly wind gusts by the observed equinoctial storms (because their momentum comes from higher altitudes where strong westerlies are present) \citep{Lucas:2014, Charnay:2015}, control the dune elongation rather than the prevailing easterlies (westward). 

This interpretation is supported by experiments done using the Titan Wind Tunnel that show that threshold wind speeds are higher than previously predicted for Titan and faster than the prevailing easterlies \citep{Burr:2015}. As with the channels observed at the Huygens landing site, it seems that many of the observed surface features may form during brief, intense periods rather than under the prevailing conditions. There is evidence that the dunes are active and therefore record the relatively recent climate conditions \citep{Barnes:2008, Neish:2010b, Lorenz:2014b, Savage:2014}.  \citet{Lorenz:2014b} estimated a reorientation timescale of $\sim$50,000 years, which is a similar timescale to climate forcing so the dunes could potentially be out of equilibrium with the present winds \citep{Lorenz:2014b, Ewing:2015}. Through an inter-GCM comparison \citet{Mcdonald:2016} showed that orbital forcing is important and must be considered when evaluating dune orientations.

In addition to dune formation, aeolian processes may be responsible for the formation of the so-called "blandlands", Titan's undifferentiated plains \citep{Lopes:2016}. Since the blandlands appear to be the most prevalent geomorphologic unit on Titan (covering $\sim$17\%), with dunes coming in a close second \citep{Lopes:2016}, an aeolian formation mechanism would mean that aeolian processes play a dominant role in shaping Titan's landscape. An analysis of material transport directions indicates that aeolian transport is the dominant transport mechanism at the equator and mid-latitudes \citep{Malaska:2016}.

\subsection{Craters} 

The dearth of craters on Titan's surface indicates that it is geologically young. The largest crater observed thus far is Menrva with a diameter of 450 km \citep{Elachi:2006}. The craters are not distributed uniformly; Xanadu is the most heavily cratered region and the equatorial dune regions and north pole have lower crater densities than average \citep{Wood:2010}. Well-preserved craters seem to be found only at low latitudes \citep{Lopes:2010}. Based on the crater size distribution, the age of the surface has been estimated to be between 200 Myr and 1 Gyr, which is quite similar to the age of Earth's surface and indicates that Titan's surface has undergone some amount of resurfacing \citep{Lorenz:2007, Neish:2012}. The craters that have been investigated in greater detail, including Sinlap and Selk craters, show evidence of fluvial and aeolian erosion \citep{LeMouelic:2008,Soderblom:2010, Wood:2010}. \citet{Neish:2013} found that craters on Titan are shallower than comparable size craters on Ganymede; they argue that the craters have been filled in from aeolian transport.

\subsection{Cryovolcanism} 

In addition to mountains (see e.g., \citet{Radebaugh:2007, Radebaugh:2011}), a number of potential cryovolcanic features have been identified through the use of RADAR and VIMS measurements. The general trend in investigation of cryovolcanic features on Titan seems to be that as more data become available for a particular feature, the less likely it seems that it is cryovolcanic. Both Ganesa Macula and Tortola Facula were initially believed to be cryovolcanic features based on RADAR \citep{Neish:2006, Lopes:2007} and VIMS measurements respectively \citep{Sotin:2005}, but data from the other instruments did not support that interpretation \citep{Hansen:2010}. Tui Regio is extremely bright at 5 $\mu$m, leading to the suggestion that it may be a region of cryovolcanic activity \citep{Barnes:2006, Barnes:2011}; measurements from RADAR may support this interpretation \citep{Mitchell:2010}, but it has been suggested that Tui Regio is actually the site of an ancient sea \citep{Moore:2010}. 

Currently, the two most probable candidate cryovolcanic features are Hotei Regio and the complex that includes Sotra Patera, Doom Mons, Mohini Fluctus, and Erebor Mons (shown in Figure \ref{fig:surface}) \citep{Lopes:2013}. Measurements from VIMS indicate that Hotei Regio is unusually bright at 5 $\mu$m \citep{Barnes:2005}, is not fluvial in origin \citep{Soderblom:2009}, and may have exhibited spectrophotometric variability in the VIMS measurements during the Cassini mission \citep{Nelson:2009}. Additionally, the flow features appear to be lobate based on RADAR mapping \citep{Wall:2009}. Others interpret the available data as an indication of a fluvial origin \citep{Moore:2011}. The Sotra Patera region consists of mountains, non-circular pits, and flow-like lobes \citep{Kirk:2010, Lopes:2013}. The large scale flow deposits known as Mohini Fluctus appear to originate from Doom Mons. Erebor Mons also has lobate deposits associated with it. Sotra Patera is 1.7 km deep, making it the deepest local depression identified thus far on Titan \citep{Lopes:2013}. The existence of cryovolcanism on Titan remains an open question, one whose answer has important implications for the origin and resupply of CH$_{4}$ in Titan's atmosphere, as discussed in Section \ref{sec:age}. 

\section{The Age and Origin of Titan's Atmosphere \label{sec:age}}

Prior to Cassini-Huygens, the origin and age of Titan's atmosphere were poorly constrained. The mostly likely sources of nitrogen in Titan's atmosphere are either N$_{2}$ or NH$_{3}$ (see e.g., \citet{Strobel:1982}), which were incorporated during formation or delivered after formation. If the original nitrogen bearing molecule was NH$_{3}$, then subsequent photolysis \citep{Lewis:1971, Atreya:1978} or reactions in Titan's interior \citep{Glein:2009, Glein:2015} are required to form the current N$_{2}$ atmosphere. The abundances of primordial noble gases are particularly important for constraining the origin of Titan's atmosphere, but could not be measured prior to Cassini-Huygens. The GCMS carried by Huygens detected $^{36}$Ar, $^{22}$Ne, and set upper limits on the abundances of Kr and Xe. The ratio of $^{36}$Ar to $^{14}$N is orders of magnitude below solar \citep{Niemann:2005, Niemann:2010} and it is therefore unlikely that large amounts of N$_{2}$ accreted during formation \citep{Barnun:1988, Niemann:2005}. The presence of radiogenic $^{40}$Ar, outgassed from Titan's rocky interior, indicates that the interior and atmosphere are connected \citep{Niemann:2005}. The non-detection of Kr and Xe is not particularly surprising; the upper limits are consistent with solar abundances and a number of other atmospheres in the solar system \citep{Owen:2009}. Additionally there are a number of possible noble gas sequestration mechanisms, both prior to and after formation: H$_{3}^{+}$ complexes \citep{Pauzat:2013}, trapping in hazes \citep{Jacovi:2008}, dissolution in lakes and seas \citep{Cordier:2010, Hodyss:2013}, and trapping in clathrate hydrates \citep{Mousis:2011, Tobie:2012}.  However, \citet{Glein:2015} recently suggested the Kr upper limit is sufficiently restrictive to indicate that Titan's CH$_{4}$ formed in Titan's interior from CO$_{2}$, rather than CH$_{4}$ clathrates that would efficiently trap Kr \citep{Niemann:2005, Atreya:2006}. Additional measurements of primordial and radiogenic noble gases are required.

\begin{figure}
{\includegraphics[width=0.9\textwidth]{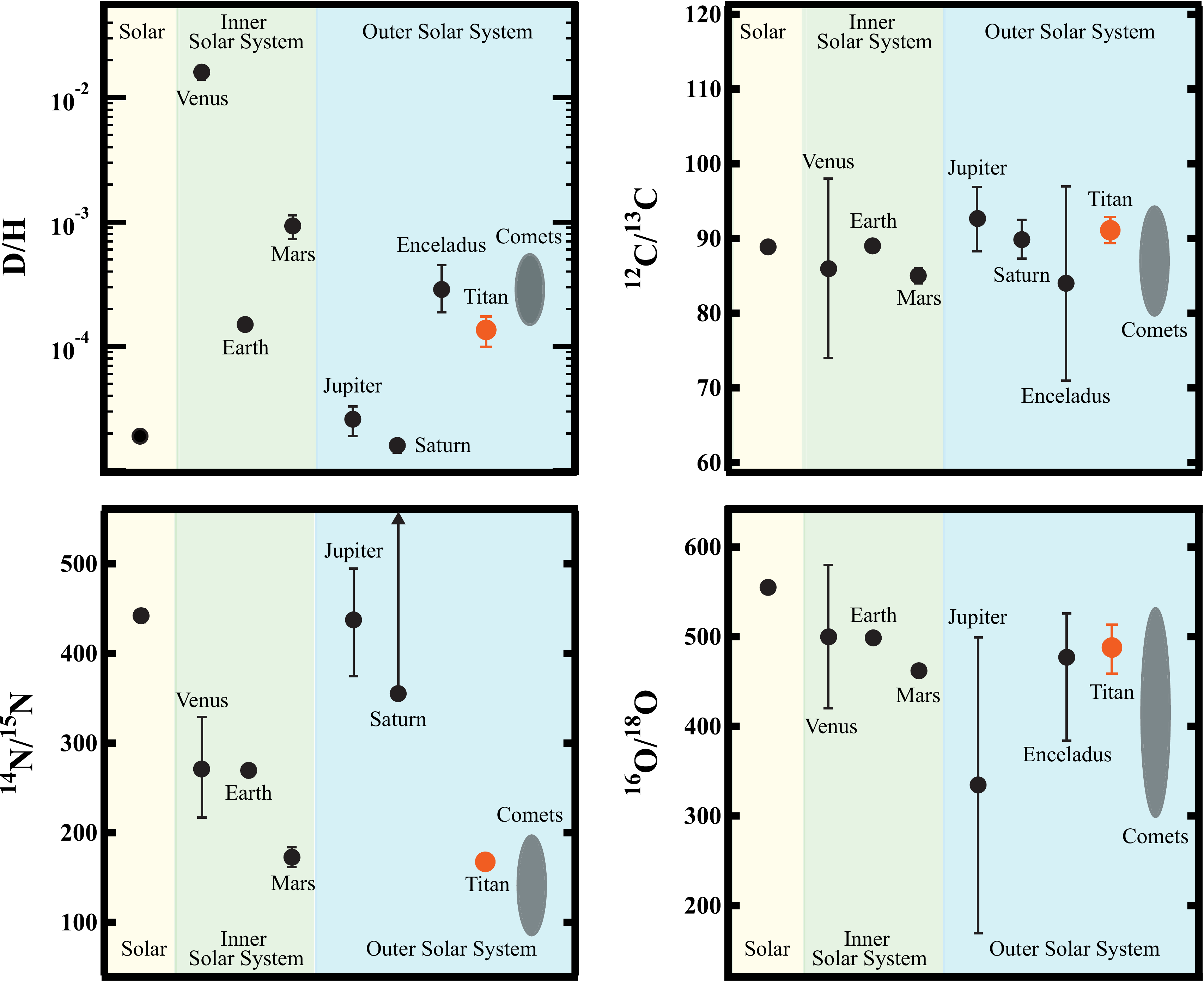}}
\caption{Titan's isotope ratios for H, C, N, and O compared with other solar system measurements. Solar- H \citep{Lodders:2003}, C \citep{Meibom:2007}, N \citep{Marty:2011}, O \citep{McKeegan:2011}, Venus- H$_{2}$O \citep{Donahue:1982}, CO$_{2}$ \citep{Bezard:1987}, N$_{2}$ \citep{Hoffman:1979}, CO$_{2}$ \citep{Hoffman:1980},  Earth- VSMOW (H, O), VPDB (C), N$_{2}$, Mars- H$_{2}$O \citep{Webster:2013}, CO$_{2}$ \citep{Mahaffy:2013}, N$_{2}$ \citep{Wong:2013}, CO$_{2}$ \citep{Mahaffy:2013}, Jupiter- H$_{2}$ \citep{Niemann:1998}, CH$_{4}$ \citep{Niemann:1998}, NH$_{3}$ \citep{Owen:2001}, H$_{2}$O \citep{Noll:1995}, Saturn- H$_{2}$ \citep{Fletcher:2009}, CH$_{4}$	 \citep{Fletcher:2009}, NH$_{3}$ \citep{Fletcher:2014}, Enceladus- H$_{2}$O, CO$_{2}$, H$_{2}$O \citep{Waite:2009}, Titan- CH$_{4}$ \citep{Niemann:2010}, N$_{2}$, \citep{Niemann:2010}, CO \citep{Serigano:2016}, Comets- H$_{2}$O \citep{BockeleeMorvan:2015}, CO$_{2}$ \citep{BockeleeMorvan:2015}, NH$_{2}$ \citep{Rousselot:2014},  H$_{2}$O \citep{BockeleeMorvan:2015}.}
\label{fig:isotopes}
\end{figure}

The $^{14}$N/$^{15}$N measurement in N$_{2}$ from GCMS (167.7$\pm$0.6, \citet{Niemann:2005, Niemann:2010}), is the lowest measured atmospheric value in the solar system and similar to a range of cometary measurements \citep{Rousselot:2014, Shinnaka:2014}. Figure \ref{fig:isotopes} places the measurements of Titan's H, C, N, and O isotopes into solar system context. Conversion of NH$_{3}$ to N$_{2}$ by photolysis \citep{Atreya:1978}, shock chemistry \citep{Mckay:1988}, and thermal decomposition in the interior \citep{Glein:2009} do not significantly alter the $^{14}$N/$^{15}$N ratio \citep{Berezhnoi:2010}. \citet{Sekine:2011} also suggested that impacts into ammonium hydrate in the crust could convert NH$_{3}$ into N$_{2}$; however, \citet{Marounina:2015} showed that impact erosion of Titan's atmosphere would be very efficient. Escape processes (Jean's, sputtering, and/or hydrodynamic) are not efficient enough to fractionate Titan's N$_{2}$ over the age of the solar system \citep{Mandt:2009, Mandt:2014}. The Ar isotopes combined with $^{14}$N/$^{15}$N indicate that Titan's nitrogen originated as ammonia from the protosolar nebula, not the subsaturnian nebula \citep{Mandt:2014}, and may have undergone conversion to N$_{2}$ in Titan's interior rather than its atmosphere \citep{Glein:2015}. Ground-based observations have set a lower limit on $^{14}$N/$^{15}$N in NH$_{3}$ in Saturn's atmosphere \citep{Fletcher:2014}, which is significantly higher than for Titan indicating that the nitrogen in the atmospheres of Titan and Saturn came from different reservoirs. 

The CH$_{4}$ in Titan's atmosphere is irreversibly destroyed by photolysis because one of the most abundant photolysis products, H$_{2}$, escapes. Photochemical models agree that the current abundance of CH$_{4}$ would be destroyed by the present day atmospheric chemistry in $\sim$30 Myr \citep{Yung:1984, Wilson:2004}, implying that the CH$_{4}$ has a continuous or episodic resupply mechanism, the models were missing important processes, and/or the atmosphere is not the age of the solar system, but we lacked information required to begin to answer these questions. Measurements of $^{12}$C/$^{13}$C in CH$_{4}$ from the INMS and CIRS agree with the measurement from GCMS \citep{Niemann:2010, Nixon:2012, Mandt:2012}, and are consistent with an assumed possible range of primordial values. Models of atmospheric escape and fractionation by chemistry (which vary in relative importance depending on the process), combined with the carbon isotope observations indicate that either no significant fractionating loss process is occurring, the methane present is very recent, or that there is a production/loss mechanism that is balancing escape over geologic timescales \citep{Nixon:2012, Mandt:2012}. Even if replenishment is occurring, the $^{12}$C/$^{13}$C still points to an atmospheric age of less than 1 Gyr.

The issue of CH$_{4}$ resupply remains an outstanding question. The lakes and seas do not represent a large enough reservoir to maintain the current atmospheric abundance over the age of the solar system. If the atmospheric methane is resupplied, methane in the subsurface ocean and/or methane clathrate hydrates in the crust are presumably the source. If the crust is methane clathrate, diffusion of C$_{2}$H$_{6}$ into the clathrate pushes CH$_{4}$ out, resupplying half of the necessary methane \citep{Choukroun:2012}. It is also possible that a substantial amount of liquid hydrocarbons is present in the subsurface, perhaps in the form of an alkanofer, which could also resupply the atmosphere. The possible detection of tropical lakes \citep{Griffith:2012b} supports the existence of a subsurface reservoir. 

Titan's D/H ratio remains poorly understood. Higher than Saturn, but lower than cometary water and Enceladus (2.9$^{+1.5}_{-0.7}\times10^{-4}$, \citet{Waite:2009}), it is difficult to explain the evolution of D/H from either of those reservoirs to Titan's present value. A measurement of D/H in cometary CH$_{4}$ and/or Titan's water ice crust would hopefully explain Titan's D/H, in the process determining if Titan's CH$_{4}$ is primordial or if it comes from conversion of CO$_{2}$ to CH$_{4}$ in Titan's interior \citep{Niemann:2010}.

Another puzzle relating to the origin of Titan's atmosphere, dating back to the Voyager era, is the abundance of carbon monoxide (CO) in Titan's atmosphere, as discussed in Section \ref{sec:chem}. The primordial CO abundance in TitanÕs atmosphere represents an additional constraint on models of Titan's formation and evolution and the thermochemical conditions where Titan's building blocks formed \citep{Tobie:2012}, which in turn constrains whether those building blocks formed in the protosolar nebula or the subsaturnian nebula. The detection of O$^{+}$ precipitating into Titan's atmosphere by the Cassini Plasma Spectrometer (CAPS) \citep{Hartle:2006}, combined with photochemical modeling of Titan's oxygen chemistry, demonstrates that Titan's oxygen has an exogenic source, Enceladus. The oxygen isotope measurements from Titan \citep{Serigano:2016} and Enceladus \citep{Waite:2009} are consistent. Given CO's extremely long lifetime in Titan's atmosphere, it is unlikely that Titan's primordial atmosphere contained significant amounts of CO \citep{Horst:2008}. This result supports the conclusion that Titan's nitrogen was originally NH$_{3}$.

\begin{figure}
 {\includegraphics[width=0.9\textwidth]{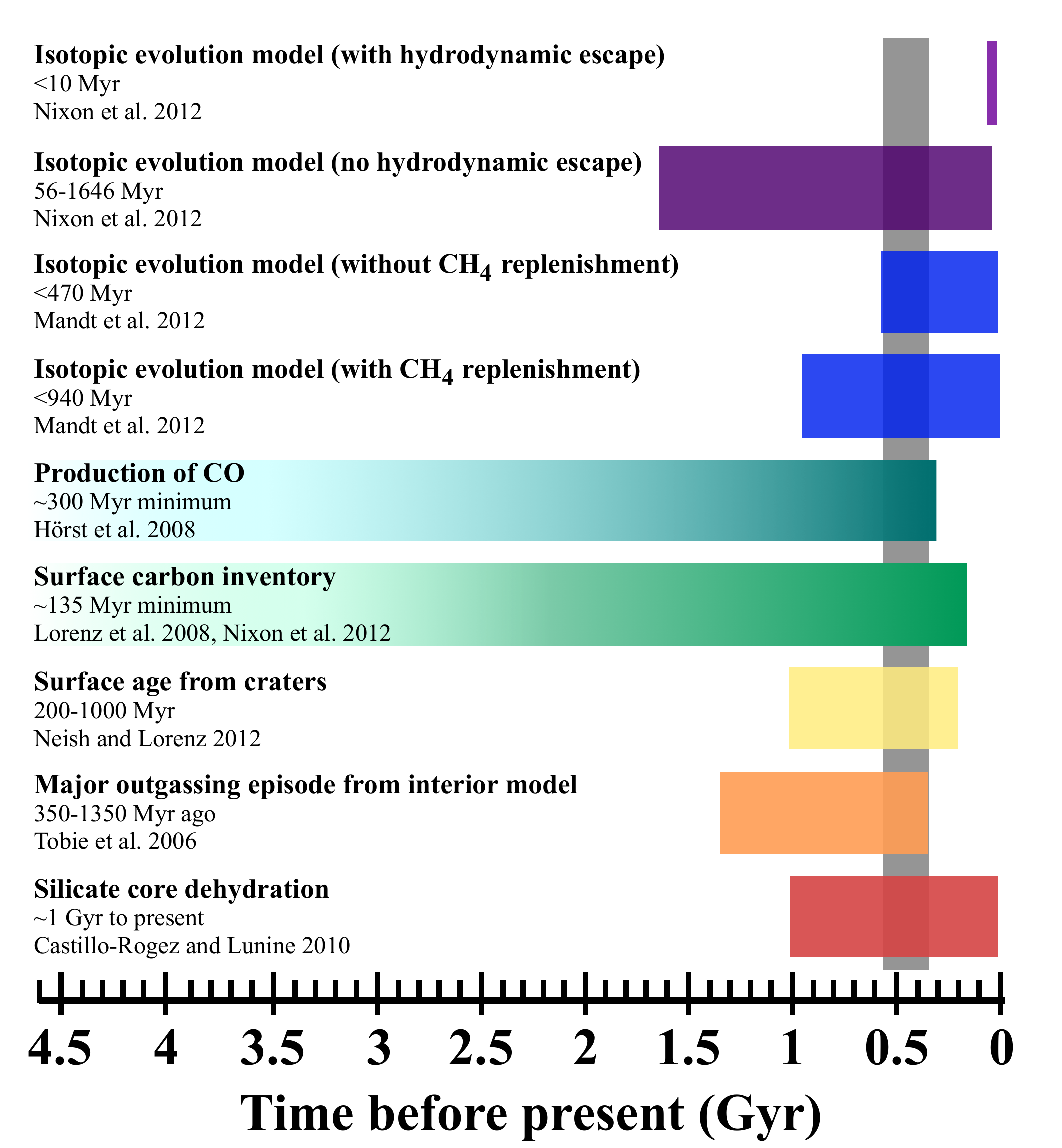}}
 \caption{Constraints on the age of Titan's atmosphere from various measurements and models \citep{Tobie:2006, Horst:2008,Lorenz:2008,Castillo:2010,Neish:2012,Nixon:2012,Mandt:2012}.}
 \label{fig:age}
 \end{figure}

There are other possible constraints on the age of Titan's atmosphere in addition to the atmospheric composition measurements and models as shown in Figure \ref{fig:age}. Interior models suggest that major outgassing episodes were likely to occur for the past 1 Gyr \citep{Tobie:2006, Castillo:2010} and the crater populations point to a surface age of $\sim$200-1000 Myr \citep{Lorenz:2007, Neish:2012}. The current observable organic inventory on the surface requires $\sim$135 Myr to accumulate \citep{Lorenz:2008b, Nixon:2012}, which is significantly less than the age of the solar system, requiring either an efficient burial mechanism or implying that the current organic deposition rates have not existed for the entire age of the solar system. Taken together, these lines of evidence make it difficult to envision a scenario in which Titan's atmosphere has existed as we see it today for the entire age of the Solar system. Additionally, overlap in these different estimates occurs around $\sim$300-500 Myr ago (without assuming CH$_{4}$ has been replenished) indicating that there is likely something fundamental about that period in Titan's history.  

There are a number of different possible scenarios if Titan's present-day atmosphere has only existed as such for $\sim$500 Myr. Many of the age constraints mentioned above and shown in Figure \ref{fig:age} constrain only the age of the CH$_{4}$ and period of active CH$_{4}$ photochemistry. It may be possible that prior to that time a different atmosphere, such as a pure N$_{2}$ atmosphere, existed. \citet{Charnay:2014} explored the behavior of a pure N$_{2}$ atmosphere, including a paleo-nitrogen cycle in which liquid nitrogen lakes and seas modified Titan's surface. This case is particularly interesting given that without a resupply mechanism, Titan's atmospheric methane will be depleted within the next 10s of Myr resulting in an N$_{2}$ atmosphere. For the future Titan case, they found that without methane, Titan's mean surface temperature would be $\sim$86.5-89.5 (depending on cloud properties) and although thin clouds would exist, the surface would be very dry \citep{Charnay:2014}.

\section{Outstanding questions and the future}

On the eve of the Voyager encounter, \citet{Caldwell:1981} wrote ``Titan is therefore also unique in being the object in the Solar System for which the atmosphere is least well understood.'' As we enter the post-Cassini-Huygens era, it is clear that Titan's atmosphere is no longer the ``least well understood.'' Nonetheless, as discussed above, a number of outstanding questions remain including:

\begin{enumerate}
\item What are the very heavy ions in the ionosphere, how do they form, and what are the implications for complexity of prebiotic chemistry?
\item What is the connection between the plumes of Enceladus and Titan's atmosphere?
\item What is the composition of the haze and how does it vary spatially and temporally?
\item How do the organic compounds produced in the atmosphere evolve once reaching the surface?
\item What are the dynamics of Titan's troposphere and how does that affect the evolution of the surface?
\item How variable is Titan's weather from year to year and how variable is the climate over longer timescales?
\item How old is Titan's current atmosphere?
\item What happened on Titan 300-500 Myrs ago?
\item Is Titan's atmospheric methane cyclic and/or episodic and if so, what are the implications of those timescales on habitability?
\item Where is the origin of Titan's methane and what is the fate of the photochemically produced ethane?
\item What is controlling Titan's H$_{2}$ profile and potential spatial variations?
\item What is the composition of the surface and on what scales is it spatially variable?
\item What is the composition of the lakes and seas and what chemistry occurs there?
\item What is the circulation in the lakes and seas and how is it affected by the atmospheric dynamics?
\item What is the composition of the dune particles and how are they produced?
\item Does cryovolcanism occur on Titan?
\end{enumerate}

Finding answers to many of these questions will require future missions to Titan; \citet{Tobie:2014} and \citet{Mitri:2014} provide excellent reviews of science goals and mission concepts for future exploration of Titan. The presence of both a subsurface water ocean \citep{Beghin:2010, Baland:2011, Bills:2011, Beghin:2012, Iess:2012} and extensive hydrocarbon seas on the surface make Titan a prime target for NASA's renewed interested in exploring ``Ocean Worlds''. Titan's diverse oceans serve as a test for the ubiquity and diversity of life \citep{Lunine:2009b}.

The wealth of data obtained by Cassini-Huygens and Earth-based observations in support of our quest to understand Titan will take many years to fully synthesize into an increased understanding of Titan's atmosphere and climate and will require additional insights from laboratory experiments and models. With the highly successful New Horizons flyby of Pluto, we now have a rich atmospheric dataset for another hazy N$_{2}$/CH$_{4}$/CO atmosphere; comparative planetology investigations in the coming years should improve our understanding of both atmospheres. Additionally, since a large number of exoplanets appear to have an aerosol absorber in their atmospheres (see e.g., \citet{Kreidberg:2014, Knutson:2014a, Knutson:2014b}), which may be photochemically produced \citep{Marley:2013}, there is increasing interest in using Titan as an exoplanet analogue \citep{Lunine:2010, Bazzon:2014, Robinson:2014}. 

Since the Voyager era, we, as a community, have made significant progress in understanding Titan's atmosphere and climate; we have transformed Titan from an enigmatic moon into a dynamic world. This transformation required the persistent and sustained effort of an international, multidisciplinary community that leveraged ground and space based observing, spacecraft measurements, laboratory experiments, and models in pursuit of one overarching goal: to understand Titan as a world. In the process, we have pushed our understanding of terrestrial processes like fluvial and aeolian erosion into completely new phase space allowing us to begin to determine the underlying fundamental physics and chemistry that drive a number of planetary processes. We unveiled a beautiful world that holds so many pieces to the puzzle of how planets form and evolve. We revealed atmospheric organic chemistry that is so complex we are forced to rethink our ideas about how atmospheres work. As discussed above, our work is not yet finished, many unanswered questions remain. Answering those questions will require a similar effort that will lead to even greater rewards.


%
%
%
%
%
%
%

\begin{acknowledgments}
This work is dedicated to my dad, Dr. William Peter Horst, who passed away unexpectedly around the time that I agreed to write this review. He still inspires me to ``slam back'' every day. I would like to thank two anonymous referees for their time and energy. The paper benefited greatly from their expertise and contributions. As a review paper, all the data shown and discussed here are in the cited references and/or in NASA's Planetary Data System (PDS) http://pds.nasa.gov.
\end{acknowledgments}

\end{article}
%
%
%
%
%
%
\clearpage

\clearpage
\begin{figure}
{\includegraphics[width=1\textwidth]{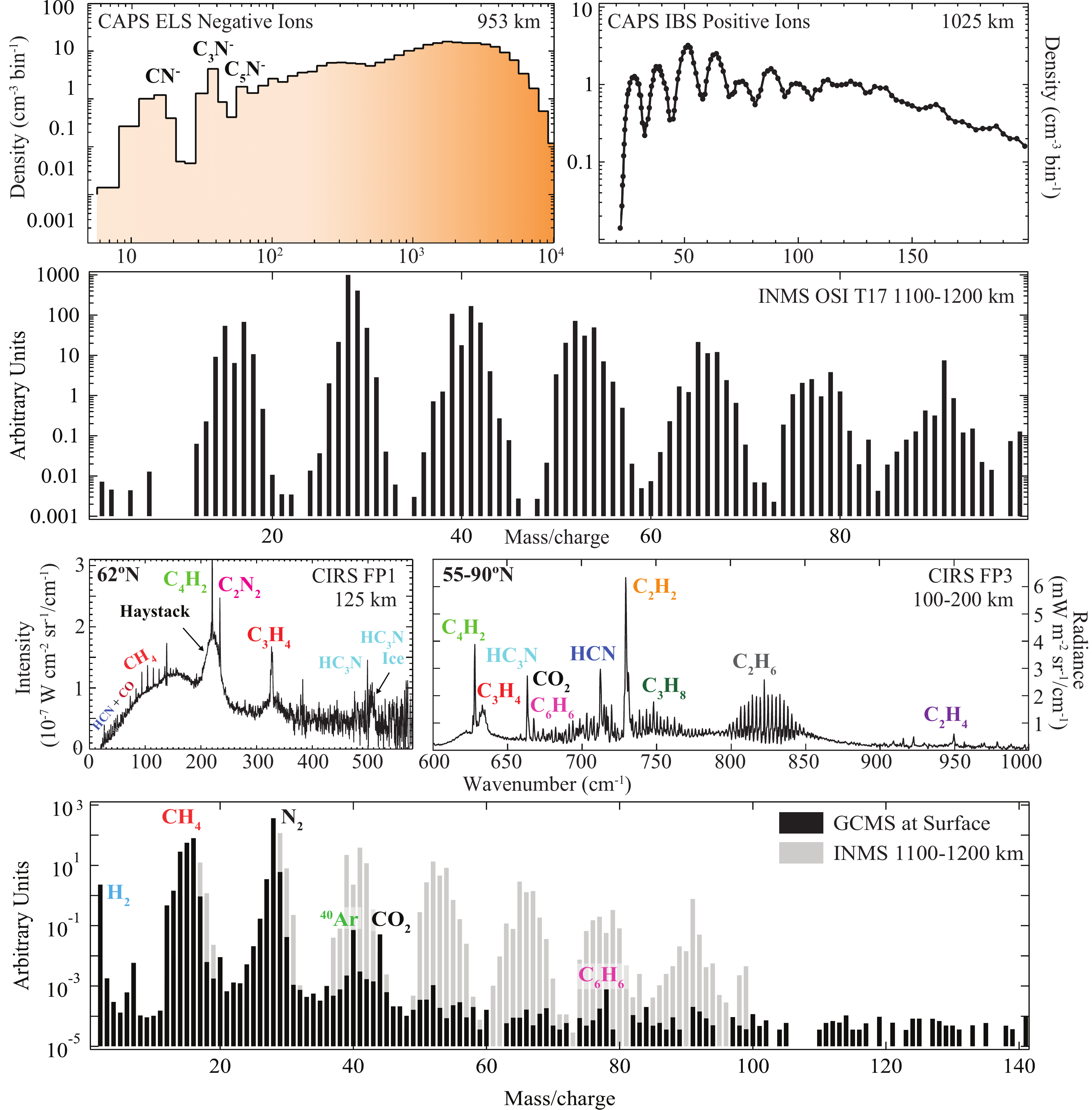}}
\label{fig:atmdata}
\end{figure}

\clearpage
\begin{figure}
\caption{The wealth of \emph{in situ} and remote sensing measurements of the composition of Titan's atmosphere from Cassini-Huygens allow us to understand the atmosphere as a system. Shown here are measurements of the thermosphere/ionosphere from CAPS ELS (top left, adapted from \citet{Coates:2007, Vuitton:2009}) and IBS (top right, adapted from \citet{Crary:2009}) and INMS. CIRS measurements of the stratosphere reveal the composition of ices and gases (FP1 data adapted from \citet{Anderson:2014}, FP3 data adapted from \citet{Bezard:2009}). Measurements from the GCMS in Titan's troposphere (bottom, adapted from \citet{Niemann:2010}) reveal that many of the trace gases produced in the thermosphere are lost to subsequent chemistry and condensation before reaching the surface.}
\label{fig:atmdata}
\end{figure}

\clearpage

\clearpage
\begin{figure}
{\includegraphics[width=1\textwidth]{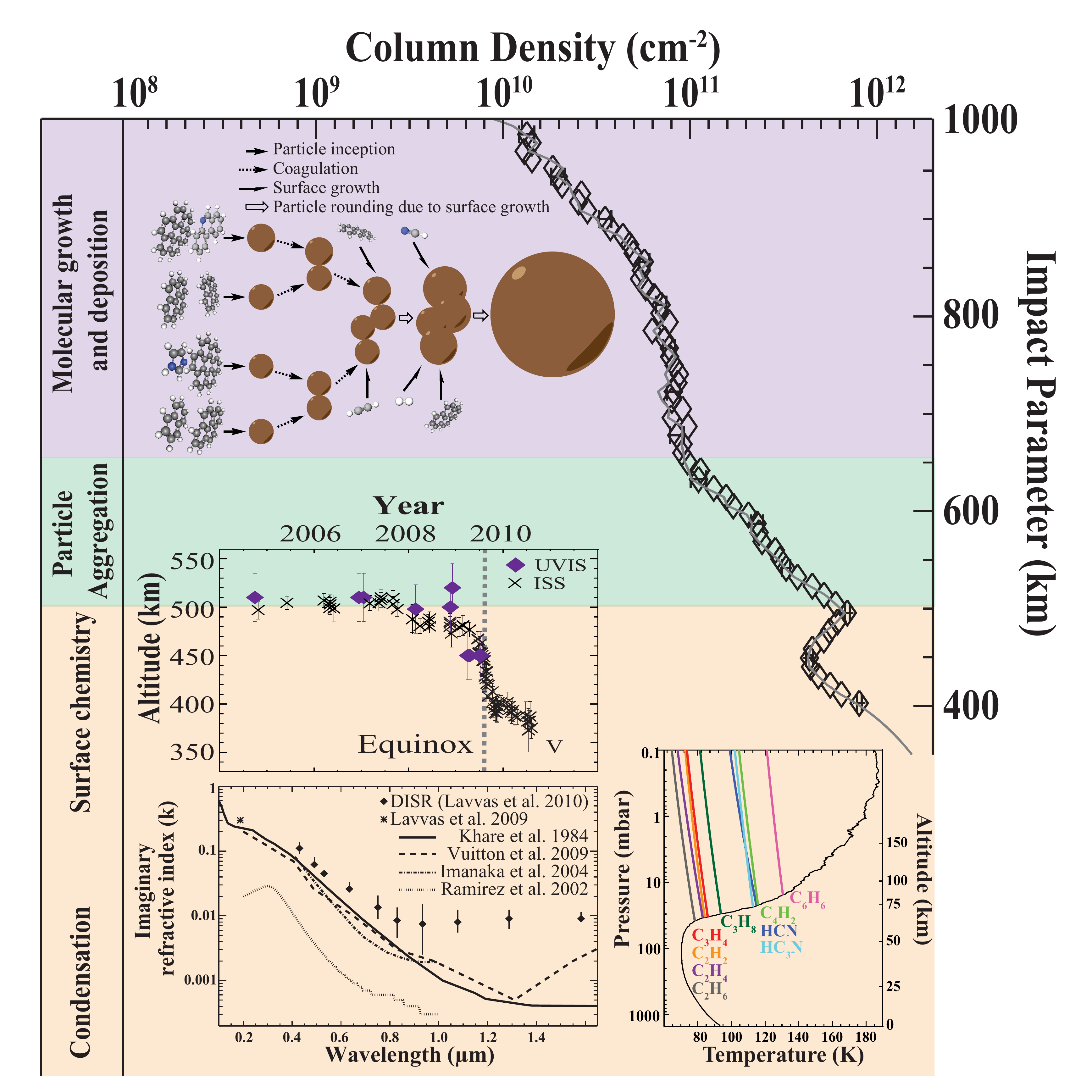}}
\label{fig:haze}
\end{figure}

\clearpage
\begin{figure}
\caption{Cassini revealed the presence of particles in Titan's atmosphere from the surface to the ionosphere. The column density from 400 to 1000 km is shown here (assuming a constant particle size, radius=12.5 nm, with altitude; particle size changes with altitude in Titan's atmosphere) and the detached haze layer is a prominent feature (adapted from \citet{Koskinen:2011}). Shown at the top is a model of particle inception and growth from \citet{Lavvas:2011}. The model divides Titan's atmosphere into 3 broad regions to explain how particles evolve as they descend through Titan's atmosphere; note that these regions refer to the dominant process in each region but the processes happen over a larger altitude range. Deep in the atmosphere, they may serve as nucleation sites for condensation and the plot at the bottom right shows where vapor pressure curves for a number of species cross the temperature profile at the Huygens landing site (adapted from \citet{Lavvas:2011c}). A comparison of DISR measurements of Titan's aerosol particles to laboratory analogues is shown bottom left (adapted from \citet{Lavvas:2010}). The detached haze layer changes altitude as a function of season as shown in the middle plot (adapted from \citet{West:2011, Koskinen:2011}).}
\label{fig:haze}
\end{figure}

\clearpage

\clearpage

\clearpage

\clearpage

\clearpage

\clearpage

\clearpage

\clearpage
\setlength{\tabcolsep}{4pt}
\begin{sidewaystable}
\begin{longtable}{lllllll}
\caption{Composition of Titan's Neutral Atmosphere \label{table:comp}}\\
\hline
&\multicolumn{4}{c}{Stratosphere}&Mesosphere&Thermosphere\\ \cline{2-4}
Formula&Ground Based$^{(a)}$&ISO$^{(f)}$/Herschel&CIRS$^{(l)}$&&UVIS$^{(r)}$&INMS (CSN)$^{(s)}$\\ \hline
\endfirsthead
\hline
Formula&Ground Based&ISO/Herschel&CIRS&&UVIS&INMS (CSN)\\ \hline
\endhead
\hline \multicolumn{3}{r}{\textit{Continued on next page}}
\endfoot
\hline
\endlastfoot 
H$_{2}$&&&9.6$\pm2.4$$\times10^{-4}$$^{(m)}$&&&3.9$\pm0.01$$\times10^{-3}$\\
$^{40}$Ar&&&&&&1.1$\pm0.03$$\times10^{-5}$\\
C$_{2}$H$_{2}$&&5.5$\pm0.5$$\times10^{-6}$&2.97$\times10^{-6}$&&5.9$\pm0.6$$\times10^{-5}$&\textit{3.1}$\mathit{\pm1.1}$$\mathit{\times10^{-4}}$\\
C$_{2}$H$_{4}$&&1.2$\pm0.3$$\times10^{-7}$&1.2$\times10^{-7}$&&1.6$\pm0.7$$\times10^{-6}$&\textit{3.1}$\mathit{\pm1.1}$$\mathit{\times10^{-4}}$\\
C$_{2}$H$_{6}$&&2.0$\pm0.8$$\times10^{-5}$&7.3$\times10^{-6}$&&&\textit{7.3}$\mathit{\pm2.6}$$\mathit{\times10^{-5}}$\\
CH$_{3}$C$_{2}$H&&1.2$\pm0.4$$\times10^{-8}$&4.8$\times10^{-9}$&&&\textit{1.4}$\mathit{\pm0.9}$$\mathit{\times10^{-4}}$\\
C$_{3}$H$_{6}$&&&2.6$\pm1.6$$\times10^{-9}$$^{(n)}$ &&&2.3$\pm0.2$$\times10^{-6}$$^{(t)}$\\ 
C$_{3}$H$_{8}$&6.2$\pm1.2$$\times10^{-7}$$^{(b)}$&2.0$\pm1.0$$\times10^{-7}$&4.5$\times10^{-7}$ &&&$\mathit{<}$\textit{4.8}$\mathit{\times10^{-5}}$\\
C$_{4}$H$_{2}$&&2.0$\pm0.5$$\times10^{-9}$&1.12$$$\times10^{-9}$&&7.6$\pm0.9$$\times10^{-7}$&\textit{6.4}$\mathit{\pm2.7}$$\mathit{\times10^{-5}}$\\
C$_{6}$H$_{6}$&&4.0$\pm3.0$$\times10^{-10}$&2.2$\times10^{-10}$ &&2.3$\pm0.3$$\times10^{-7}$&8.95$\pm0.44$$\times10^{-7}$\\
HCN	&5$\times10^{-7}$&3.0$\pm0.5$$\times10^{-7}$&6.7$\times10^{-8}$&&1.6$\pm0.7$$\times10^{-5}$&\\
HNC&4.9$\pm0.3$$\times10^{-9}$$^{(c)}$&4.5$\pm1.2$$\times10^{-9}$$^{(g)}$&&&&\\
HC$_{3}$N&3$\times10^{-11}$&5.0$\pm3.5$$\times10^{-10}$&2.8$\times10^{-10}$&& 2.4$\pm0.3$$\times10^{-6}$&\textit{3.2}$\mathit{\pm0.7}$$\mathit{\times10^{-5}}$\\
CH$_{3}$CN&8$\times10^{-9}$&&$<$1.1$\times10^{-7}$$^{(o)}$&&&\textit{3.1}$\mathit{\pm0.7}$$\mathit{\times10^{-5}}$\\ 
C$_{2}$H$_{5}$CN&2.8$\times10^{-10}$$^{(d)}$&&&&&\\
C$_{2}$N$_{2}$&&&9$\times10^{-10}$$^{(p)}$&&&\textit{4.8}$\mathit{\pm0.8}$$\mathit{\times10^{-5}}$\\
NH$_{3}$	&&$<$1.9$\times10^{-10}$$^{(h)}$&$<$1.3$\times10^{-9}$$^{(o)}$&&&2.99$\pm0.22$$\times10^{-5}$\\
CO&5.1$\pm0.4$$\times10^{-5}$$^{(e)}$&$4.0\pm5\times10^{-5}$$^{(i)}$&4.7$\pm0.8$$\times10^{-5}$$^{(q)}$ &&&\\
H$_{2}$O	&&8$\times10^{-9}$$^{(j)}$/7$\times10^{-10}$$^{(k)}$&4.5$\pm1.5$$\times10^{-10}$$^{(q)}$&&&$<$3.42$\times10^{-6}$\\
CO$_{2}$	&&2.0$\pm0.2$$\times10^{-8}$&1.1$\times10^{-8}$&&&$<$8.49$\times10^{-7}$\\
\hline
\multicolumn{7}{l}{All values come from the source and work referenced in the table header unless otherwise noted in the table.}\\
\multicolumn{7}{l}{$^{(a)}$\citet{Marten:2002} (at 200 km) $^{(b)}$\citet{Roe:2003} $^{(c)}$\citet{Cordiner:2014} (constant profile at 400 km)}\\ 
\multicolumn{7}{l}{$^{(d)}$\citet{Cordiner:2015} (gradient profile at 200 km) $^{(e)}$\citet{Gurwell:2004} $^{(f)}$\citet{Coustenis:2003}}\\ 
\multicolumn{7}{l}{$^{(g)}$\citet{Moreno:2011} (constant profile at 400 km) $^{(h)}$\citet{Teanby:2013} (3-$\sigma$ upper limit, peak sensitivity}\\
\multicolumn{7}{l}{at 75 km) $^{(i)}$\citet{Courtin:2011} $^{(j)}$\citet{Coustenis:1998} (at 400 km) $^{(k)}$\citet{Moreno:2012} (from}\\
\multicolumn{7}{l}{the S$_{a}$ profile at 400 km) $^{(l)}$\citet{Coustenis:2010} (at 5$^{\circ}$S; HCN, C$_{6}$H$_{6}$, C$_{3}$H$_{8}$, C$_{4}$H$_{2}$, and HC$_{3}$N}\\
\multicolumn{7}{l}{exhibit latitudinal variations \citep{Coustenis:2007}), values averaged over TB-T44 assuming constant}\\
\multicolumn{7}{l}{vertical profiles, for vertical variations see \citet{Vinatier:2010}) $^{(m)}$\citet{Courtin:2007}}\\
\multicolumn{7}{l}{$^{(o)}$\citet{Nixon:2010} (3-$\sigma$ upper limits at 25$^{\circ}$S) $^{(p)}$\citet{Teanby:2006} (3-$\sigma$ upper limit at 50$^{\circ}$N)}\\
\multicolumn{7}{l}{$^{(n)}$\citet{Nixon:2013b} (at 225 km) $^{(q)}$\citet{Cottini:2012} (using constant VMR at 230 km)}\\
\multicolumn{7}{l}{$^{(r)}$\citet{Koskinen:2011} (at $\sim$600 km) $^{(s)}$\citet{Cui:2009} (at 1077 km from T5, 16, 18, 19, 21, 23, 25-30,}\\
\multicolumn{7}{l}{32, 36, 37 using INMS CSN mode, \textit{italics} indicates corrected values, signals from C$_{2}$H$_{2}$ and C$_{2}$H$_{4}$ are difficult to}\\
\multicolumn{7}{l}{separate so the reported value is for both species combined) $^{(t)}$\citet{Magee:2009} (global average at 1050 km) }\\
\end{longtable}
\end{sidewaystable}

\end{document}